%
%
%
\def\unredoffs{} \def\redoffs{\voffset=-.31truein\hoffset=-.48truein}
\def\speclscape{}
%
%
%
%
%
\newbox\leftpage \newdimen\fullhsize \newdimen\hstitle \newdimen\hsbody
\tolerance=1000\hfuzz=2pt
\catcode`\@=11 
\ifx\hyperdef\UNd@FiNeD\def\hyperdef#1#2#3#4{#4}\def\hyperref#1#2#3#4{#4}\fi
\def\bigans{b }
\def\answ{b }
%
\ifx\answ\bigans\message{(This will come out unreduced.}
\magnification=1200\unredoffs\baselineskip=16pt plus 2pt minus 1pt
\hsbody=\hsize \hstitle=\hsize 
\else\message{(This will be reduced.} \let\l@r=L
\magnification=1000\baselineskip=16pt plus 2pt minus 1pt \vsize=7truein
\redoffs \hstitle=8truein\hsbody=4.75truein\fullhsize=10truein\hsize=\hsbody
\output={\ifnum\pageno=0 
  \shipout\vbox{\speclscape{\hsize\fullhsize\makeheadline}
    \hbox to \fullhsize{\hfill\pagebody\hfill}}\advancepageno
  \else
  \almostshipout{\leftline{\vbox{\pagebody\makefootline}}}\advancepageno
  \fi}
\def\almostshipout#1{\if L\l@r \count1=1 \message{[\the\count0.\the\count1]}
      \global\setbox\leftpage=#1 \global\let\l@r=R
 \else \count1=2
  \shipout\vbox{\speclscape{\hsize\fullhsize\makeheadline}
      \hbox to\fullhsize{\box\leftpage\hfil#1}}  \global\let\l@r=L\fi}
\fi
%
\newcount\yearltd\yearltd=\year\advance\yearltd by -2000

\def\Title#1#2{\nopagenumbers\abstractfont\hsize=\hstitle\rightline{#1}%
\vskip 1in\centerline{\titlefont #2}\abstractfont\vskip .5in\pageno=0}
\def\Date#1{\vfill\leftline{#1}\tenpoint\supereject\global\hsize=\hsbody%
\footline={\hss\tenrm\hyperdef\hypernoname{page}\folio\folio\hss}}%
%

\def\draftmode{\message{ DRAFTMODE }\def\draftdate{{\rm preliminary draft:
\number\month/\number\day/\number\yearltd\ \ \hourmin}}%
\headline={\hfil\draftdate}\writelabels\baselineskip=20pt plus 2pt minus 2pt
 {\count255=\time\divide\count255 by 60 \xdef\hourmin{\number\count255}
  \multiply\count255 by-60\advance\count255 by\time
  \xdef\hourmin{\hourmin:\ifnum\count255<10 0\fi\the\count255}}}
\def\nolabels{\def\wrlabeL##1{}\def\eqlabeL##1{}\def\reflabeL##1{}}
\def\writelabels{\def\wrlabeL##1{\leavevmode\vadjust{\rlap{\smash%
{\line{{\escapechar=` \hfill\rlap{\sevenrm\hskip.03in\string##1}}}}}}}%
\def\eqlabeL##1{{\escapechar-1\rlap{\sevenrm\hskip.05in\string##1}}}%
\def\reflabeL##1{\noexpand\llap{\noexpand\sevenrm\string\string\string##1}}}
\nolabels
%
\global\newcount\secno \global\secno=0
\global\newcount\meqno \global\meqno=1
\def\s@csym{}
\def\newsec#1{\global\advance\secno by1%
{\toks0{#1}\message{(\the\secno. \the\toks0)}}%
\global\subsecno=0\eqnres@t\let\s@csym\secsym\xdef\secn@m{\the\secno}\noindent
{\bf\hyperdef\hypernoname{section}{\the\secno}{\the\secno.} #1}%
\writetoca{{\string\hyperref{}{section}{\the\secno}{\it\the\secno.}} {{\it #1} }}%
\par\nobreak\medskip\nobreak}
\def\eqnres@t{\xdef\secsym{\the\secno.}\global\meqno=1\bigbreak\bigskip}
\def\sequentialequations{\def\eqnres@t{\bigbreak}}\xdef\secsym{}
\global\newcount\subsecno \global\subsecno=0
\def\subsec#1{\global\advance\subsecno by1%
{\toks0{#1}\message{(\s@csym\the\subsecno. \the\toks0)}}%
\ifnum\lastpenalty>9000\else\bigbreak\fi       \global\subsubsecno=0
\noindent{\it\hyperdef\hypernoname{subsection}{\secn@m.\the\subsecno}%
{\secn@m.\the\subsecno.} #1}\writetoca{\string\quad
{\string\hyperref{}{subsection}{\secn@m.\the\subsecno}{\secn@m.\the\subsecno.}}
{#1}}\par\nobreak\medskip\nobreak}
\def\appendix#1#2{\global\meqno=1\global\subsecno=0\xdef\secsym{\hbox{#1.}}%
\bigbreak\bigskip\noindent{\bf Appendix \hyperdef\hypernoname{appendix}{#1}%
{#1.} #2}{\toks0{(#1. #2)}\message{\the\toks0}}%
\xdef\s@csym{#1.}\xdef\secn@m{#1}%
\writetoca{\string\hyperref{}{appendix}{#1}{{\it Appendix} {\it #1.}} {\it #2}}%
\par\nobreak\medskip\nobreak}
%
%
\def\checkm@de#1#2{\ifmmode{\def\f@rst##1{##1}\hyperdef\hypernoname{equation}%
{#1}{#2}}\else\hyperref{}{equation}{#1}{#2}\fi}
\def\eqnn#1{\DefWarn#1\xdef #1{(\noexpand\relax\noexpand\checkm@de%
{\s@csym\the\meqno}{\secsym\the\meqno})}%
\wrlabeL#1\writedef{#1\leftbracket#1}\global\advance\meqno by1}
\def\f@rst#1{\c@t#1a\em@ark}\def\c@t#1#2\em@ark{#1}
\def\eqna#1{\DefWarn#1\wrlabeL{#1$\{\}$}%
\xdef #1##1{(\noexpand\relax\noexpand\checkm@de%
{\s@csym\the\meqno\noexpand\f@rst{##1}}{\hbox{$\secsym\the\meqno##1$}})}
\writedef{#1\numbersign1\leftbracket#1{\numbersign1}}\global\advance\meqno by1}
\def\eqn#1#2{\DefWarn#1%
\xdef #1{(\noexpand\hyperref{}{equation}{\s@csym\the\meqno}%
{\secsym\the\meqno})}$$#2\eqno(\hyperdef\hypernoname{equation}%
{\s@csym\the\meqno}{\secsym\the\meqno})\eqlabeL#1$$%
\writedef{#1\leftbracket#1}\global\advance\meqno by1}
\def\xeqn{\expandafter\xe@n}\def\xe@n(#1){#1}
\def\xeqna#1{\expandafter\xe@n#1}
\def\eqns#1{(\e@ns #1{\hbox{}})}
\def\e@ns#1{\ifx\UNd@FiNeD#1\message{eqnlabel \string#1 is undefined.}%
\xdef#1{(?.?)}\fi{\let\hyperref=\relax\xdef\next{#1}}%
\ifx\next\em@rk\def\next{}\else%
\ifx\next#1\xeqn#1\else\def\n@xt{#1}\ifx\n@xt\next#1\else\xeqna#1\fi
\fi\let\next=\e@ns\fi\next}

\def\DefWarn#1{\ifx\UNd@FiNeD#1\else
\immediate\write16{*** WARNING: the label \string#1 is already defined ***}\fi}
%
\newskip\footskip\footskip14pt plus 1pt minus 1pt 
\def\footnotefont{\ninepoint}\def\f@t#1{\footnotefont #1\@foot}
\def\f@@t{\baselineskip\footskip\bgroup\footnotefont\aftergroup\@foot\let\next}
\setbox\strutbox=\hbox{\vrule height9.5pt depth4.5pt width0pt}
\global\newcount\ftno \global\ftno=0
\def\foot{\global\advance\ftno by1\def\foot@rg{\hyperref{}{footnote}%
{\the\ftno}{\the\ftno}\xdef\foot@rg{\noexpand\hyperdef\noexpand\hypernoname%
{footnote}{\the\ftno}{\the\ftno}}}\footnote{$^{\foot@rg}$}}
%
\newwrite\ftfile
\def\footend{\def\foot{\global\advance\ftno by1\chardef\wfile=\ftfile
\hyperref{}{footnote}{\the\ftno}{$^{\the\ftno}$}%
\ifnum\ftno=1\immediate\openout\ftfile=\jobname.fts\fi%
\immediate\write\ftfile{\noexpand\smallskip%
\noexpand\item{\noexpand\hyperdef\noexpand\hypernoname{footnote}
{\the\ftno}{f\the\ftno}:\ }\pctsign}\findarg}%
\def\footatend{\vfill\eject\immediate\closeout\ftfile{\parindent=20pt
\centerline{\bf Footnotes}\nobreak\bigskip\input \jobname.fts }}}
\def\footatend{}
%
%
\global\newcount\refno \global\refno=1
\newwrite\rfile
\def\ref{[\hyperref{}{reference}{\the\refno}{\the\refno}]\nref}
\def\nref#1{\DefWarn#1%
\xdef#1{[\noexpand\hyperref{}{reference}{\the\refno}{\the\refno}]}%
\writedef{#1\leftbracket#1}%
\ifnum\refno=1\immediate\openout\rfile=\jobname.refs\fi
\chardef\wfile=\rfile\immediate\write\rfile{\noexpand\item{[\noexpand\hyperdef%
\noexpand\hypernoname{reference}{\the\refno}{\the\refno}]\ }%
\reflabeL{#1\hskip.31in}\pctsign}\global\advance\refno by1\findarg}
\def\findarg#1#{\begingroup\obeylines\newlinechar=`\^^M\pass@rg}
{\obeylines\gdef\pass@rg#1{\writ@line\relax #1^^M\hbox{}^^M}%
\gdef\writ@line#1^^M{\expandafter\toks0\expandafter{\striprel@x #1}%
\edef\next{\the\toks0}\ifx\next\em@rk\let\next=\endgroup\else\ifx\next\empty%
\else\immediate\write\wfile{\the\toks0}\fi\let\next=\writ@line\fi\next\relax}}
\def\striprel@x#1{} \def\em@rk{\hbox{}}
\def\lref{\begingroup\obeylines\lr@f}
\def\lr@f#1#2{\DefWarn#1\gdef#1{\let#1=\UNd@FiNeD\ref#1{#2}}\endgroup\unskip}

\def\addref#1{\immediate\write\rfile{\noexpand\item{}#1}} 
\def\listrefs{\footatend\vfill\supereject\immediate\closeout\rfile\writestoppt
\baselineskip=\footskip\centerline{{\bf References}}\bigskip{\parindent=20pt%
\frenchspacing\escapechar=` \input \jobname.refs\vfill\eject}\nonfrenchspacing}
\def\startrefs#1{\immediate\openout\rfile=\jobname.refs\refno=#1}
\def\xref{\expandafter\xr@f}\def\xr@f[#1]{#1}
\def\refs#1{\count255=1[\r@fs #1{\hbox{}}]}
\def\r@fs#1{\ifx\UNd@FiNeD#1\message{reflabel \string#1 is undefined.}%
\nref#1{need to supply reference \string#1.}\fi%
\vphantom{\hphantom{#1}}{\let\hyperref=\relax\xdef\next{#1}}%
\ifx\next\em@rk\def\next{}%
\else\ifx\next#1\ifodd\count255\relax\xref#1\count255=0\fi%
\else#1\count255=1\fi\let\next=\r@fs\fi\next}
%

%
\newwrite\ffile\global\newcount\figno \global\figno=1
\def\fig{fig.~\hyperref{}{figure}{\the\figno}{\the\figno}\nfig}
\def\nfig#1{\DefWarn#1%
\xdef#1{fig.~\noexpand\hyperref{}{figure}{\the\figno}{\the\figno}}%
\writedef{#1\leftbracket fig.\noexpand~\xfig#1}%
\ifnum\figno=1\immediate\openout\ffile=\jobname.figs\fi\chardef\wfile=\ffile%
{\let\hyperref=\relax
\immediate\write\ffile{\noexpand\medskip\noexpand\item{Fig.\ %
\noexpand\hyperdef\noexpand\hypernoname{figure}{\the\figno}{\the\figno}. }
\reflabeL{#1\hskip.55in}\pctsign}}\global\advance\figno by1\findarg}
\def\listfigs{\vfill\eject\immediate\closeout\ffile{\parindent40pt
\baselineskip14pt\centerline{{\bf Figure Captions}}\nobreak\medskip
\escapechar=` \input \jobname.figs\vfill\eject}}
\def\xfig{\expandafter\xf@g}\def\xf@g fig.\penalty\@M\ {}
\def\figs#1{figs.~\f@gs #1{\hbox{}}}
\def\f@gs#1{{\let\hyperref=\relax\xdef\next{#1}}\ifx\next\em@rk\def\next{}\else
\ifx\next#1\xfig #1\else#1\fi\let\next=\f@gs\fi\next}
\def\figin{\epsfcheck\figin}\def\figins{\epsfcheck\figins}
\def\epsfcheck{\ifx\epsfbox\UNd@FiNeD
\message{(NO epsf.tex, FIGURES WILL BE IGNORED)}
\gdef\figin##1{\vskip2in}\gdef\figins##1{\hskip.5in}
\else\message{(FIGURES WILL BE INCLUDED)}%
\gdef\figin##1{##1}\gdef\figins##1{##1}\fi}
\def\DefWarn#1{}
\def\figinsert{\goodbreak\midinsert}
\def\ifig#1#2#3{\DefWarn#1\xdef#1{Fig.~\noexpand\hyperref{}{figure}%
{\the\figno}{\the\figno}}\writedef{#1\leftbracket fig.\noexpand~\xfig#1}%
\figinsert\figin{\centerline{#3}}\medskip\centerline{\vbox{\baselineskip12pt
\advance\hsize by -1truein\noindent\wrlabeL{#1=#1}\footnotefont%
{\bf Fig.~\hyperdef\hypernoname{figure}{\the\figno}{\the\figno}:} #2}}
\bigskip\endinsert\global\advance\figno by1}
\newwrite\lfile
{\escapechar-1\xdef\pctsign{\string\%}\xdef\leftbracket{\string\{}
\xdef\rightbracket{\string\}}\xdef\numbersign{\string\#}}
\def\writedefs{\immediate\openout\lfile=\jobname.defs \def\writedef##1{%
{\let\hyperref=\relax\let\hyperdef=\relax\let\hypernoname=\relax
 \immediate\write\lfile{\string\def\string##1\rightbracket}}}}%
\def\writestop{\def\writestoppt{\immediate\write\lfile{\string\pageno
 \the\pageno\string\startrefs\leftbracket\the\refno\rightbracket
 \string\def\string\secsym\leftbracket\secsym\rightbracket
 \string\secno\the\secno\string\meqno\the\meqno}\immediate\closeout\lfile}}
\def\writestoppt{}\def\writedef#1{}
\def\seclab#1{\DefWarn#1%
\xdef #1{\noexpand\hyperref{}{section}{\the\secno}{\the\secno}}%
\writedef{#1\leftbracket#1}\wrlabeL{#1=#1}}
\def\subseclab#1{\DefWarn#1%
\xdef #1{\noexpand\hyperref{}{subsection}{\secn@m.\the\subsecno}%
{\secn@m.\the\subsecno}}\writedef{#1\leftbracket#1}\wrlabeL{#1=#1}}
\def\applab#1{\DefWarn#1%
\xdef #1{\noexpand\hyperref{}{appendix}{\secn@m}{\secn@m}}%
\writedef{#1\leftbracket#1}\wrlabeL{#1=#1}}
\newwrite\tfile \def\writetoca#1{}
\def\leaderfill{\leaders\hbox to 1em{\hss.\hss}\hfill}
\def\writetoc{\immediate\openout\tfile=\jobname.toc
   \def\writetoca##1{{\edef\next{\write\tfile{\noindent ##1
   \string\leaderfill {\string\hyperref{}{page}{\noexpand\number\pageno}%
                       {\noexpand\number\pageno}} \par}}\next}}}
\newread\ch@ckfile
\def\listtoc{\immediate\closeout\tfile\immediate\openin\ch@ckfile=\jobname.toc
\ifeof\ch@ckfile\message{no file \jobname.toc, no table of contents this pass}%
\else\closein\ch@ckfile\centerline{\bf Contents}\nobreak\medskip%
{\baselineskip=18.5pt  \footnotefont
\parskip=2pt\catcode`\@=12\input\jobname.toc
\catcode`\@=12\bigbreak\bigskip}\fi}
\catcode`\@=12 
%
\edef\tfontsize{\ifx\answ\bigans scaled\magstep3\else scaled\magstep4\fi}
\font\titlerm=cmr10 \tfontsize \font\titlerms=cmr7 \tfontsize
\font\titlermss=cmr5 \tfontsize \font\titlei=cmmi10 \tfontsize
\font\titleis=cmmi7 \tfontsize \font\titleiss=cmmi5 \tfontsize
\font\titlesy=cmsy10 \tfontsize \font\titlesys=cmsy7 \tfontsize
\font\titlesyss=cmsy5 \tfontsize \font\titleit=cmti10 \tfontsize
\skewchar\titlei='177 \skewchar\titleis='177 \skewchar\titleiss='177
\skewchar\titlesy='60 \skewchar\titlesys='60 \skewchar\titlesyss='60
\def\titlefont{\def\rm{\fam0\titlerm}
\textfont0=\titlerm \scriptfont0=\titlerms \scriptscriptfont0=\titlermss
\textfont1=\titlei \scriptfont1=\titleis \scriptscriptfont1=\titleiss
\textfont2=\titlesy \scriptfont2=\titlesys \scriptscriptfont2=\titlesyss
\textfont\itfam=\titleit \def\it{\fam\itfam\titleit}\rm}
 \ifx\answ\bigans\else scaled\magstep1\fi
\ifx\answ\bigans\def\abstractfont{\tenpoint}\else
\font\absit=cmti10 scaled \magstep1
\font\abssl=cmsl10 scaled \magstep1
\font\absrm=cmr10 scaled\magstep1 \font\absrms=cmr7 scaled\magstep1
\font\absrmss=cmr5 scaled\magstep1 \font\absi=cmmi10 scaled\magstep1
\font\absis=cmmi7 scaled\magstep1 \font\absiss=cmmi5 scaled\magstep1
\font\abssy=cmsy10 scaled\magstep1 \font\abssys=cmsy7 scaled\magstep1
\font\abssyss=cmsy5 scaled\magstep1 \font\absbf=cmbx10 scaled\magstep1
\skewchar\absi='177 \skewchar\absis='177 \skewchar\absiss='177
\skewchar\abssy='60 \skewchar\abssys='60 \skewchar\abssyss='60
\def\abstractfont{\def\rm{\fam0\absrm}
\textfont0=\absrm \scriptfont0=\absrms \scriptscriptfont0=\absrmss
\textfont1=\absi \scriptfont1=\absis \scriptscriptfont1=\absiss
\textfont2=\abssy \scriptfont2=\abssys \scriptscriptfont2=\abssyss
\textfont\itfam=\absit \def\it{\fam\itfam\absit}\def\footnotefont{\tenpoint}%
\textfont\slfam=\abssl \def\sl{\fam\slfam\abssl}%
\textfont\bffam=\absbf \def\bf{\fam\bffam\absbf}\rm}\fi
\def\tenpoint{\def\rm{\fam0\tenrm}
\textfont0=\tenrm \scriptfont0=\sevenrm \scriptscriptfont0=\fiverm
\textfont1=\teni  \scriptfont1=\seveni  \scriptscriptfont1=\fivei
\textfont2=\tensy \scriptfont2=\sevensy \scriptscriptfont2=\fivesy
\textfont\itfam=\tenit \def\it{\fam\itfam\tenit}\def\footnotefont{\ninepoint}%
\textfont\bffam=\tenbf \def\bf{\fam\bffam\tenbf}\def\sl{\fam\slfam\tensl}\rm}
\font\ninerm=cmr9 \font\sixrm=cmr6 \font\ninei=cmmi9 \font\sixi=cmmi6
\font\ninesy=cmsy9 \font\sixsy=cmsy6 \font\ninebf=cmbx9
\font\nineit=cmti9 \font\ninesl=cmsl9 \skewchar\ninei='177
\skewchar\sixi='177 \skewchar\ninesy='60 \skewchar\sixsy='60
\def\ninepoint{\def\rm{\fam0\ninerm}
\textfont0=\ninerm \scriptfont0=\sixrm \scriptscriptfont0=\fiverm
\textfont1=\ninei \scriptfont1=\sixi \scriptscriptfont1=\fivei
\textfont2=\ninesy \scriptfont2=\sixsy \scriptscriptfont2=\fivesy
\textfont\itfam=\ninei \def\it{\fam\itfam\nineit}\def\sl{\fam\slfam\ninesl}%
\textfont\bffam=\ninebf \def\bf{\fam\bffam\ninebf}\rm}
%
%
\def\noblackbox{\overfullrule=0pt}
\hyphenation{anom-aly anom-alies coun-ter-term coun-ter-terms}
\def\inv{^{\raise.15ex\hbox{${\scriptscriptstyle -}$}\kern-.05em 1}}

\def\Dsl{\,\raise.15ex\hbox{/}\mkern-13.5mu D} 
\def\dsl{\raise.15ex\hbox{/}\kern-.57em\partial}

\def\lspace{\ifx\answ\bigans{}\else\qquad\fi}
\def\lbspace{\ifx\answ\bigans{}\else\hskip-.2in\fi} 
\def\boxeqn#1{\vcenter{\vbox{\hrule\hbox{\vrule\kern3pt\vbox{\kern3pt
	\hbox{${\displaystyle #1}$}\kern3pt}\kern3pt\vrule}\hrule}}}
\def\mbox#1#2{\vcenter{\hrule \hbox{\vrule height#2in
		\kern#1in \vrule} \hrule}}  
%

\def\darr#1{\raise1.5ex\hbox{$\leftrightarrow$}\mkern-16.5mu #1}

\def\half{{\textstyle{1\over2}}} 
\def\roughly#1{\raise.3ex\hbox{$#1$\kern-.75em\lower1ex\hbox{$\sim$}}}

\global\newcount\subsubsecno \global\subsubsecno=0
\def\subsubsec#1{\global\advance\subsubsecno by1%
{\toks0{#1}\message{(\the\secno\the\subsecno\the\subsubsecno. \the\toks0)}}%
\ifnum\lastpenalty>9000\else\bigbreak\fi
\noindent{\it\hyperdef\hypernoname{subsubsection}{\the\secno.\the\subsecno\the\subsubsecno}%
{\the\secno.\the\subsecno.\the\subsubsecno.} #1}
\par\nobreak\medskip\nobreak}
\def\boxit#1{\vbox{\hrule\hbox{\vrule\kern8pt
\vbox{\hbox{\kern8pt}\hbox{\vbox{#1}}\hbox{\kern8pt}}
\kern8pt\vrule}\hrule}}
\def\mathboxit#1{\vbox{\hrule\hbox{\vrule\kern8pt\vbox{\kern8pt
\hbox{$\displaystyle #1$}\kern8pt}\kern8pt\vrule}\hrule}}
\def\slashchar#1{\setbox0=\hbox{$#1$}           
   \dimen0=\wd0                                 
   \setbox1=\hbox{/} \dimen1=\wd1               
   \ifdim\dimen0>\dimen1                        
      \rlap{\hbox to \dimen0{\hfil/\hfil}}      
      #1                                        
   \else                                        
      \rlap{\hbox to \dimen1{\hfil$#1$\hfil}}   
      /                                         
   \fi}
\def\sqr#1#2{{\vcenter{\vbox{\hrule height.#2pt
         \hbox{\vrule width.#2pt height#1pt \kern#1pt
            \vrule width.#2pt}
         \hrule height.#2pt}}}}


\input amssym.def
\input amssym.tex
\noblackbox
\baselineskip=14.5pt

\def\comment#1{{}}
\def\ss#1{{\scriptstyle{#1}}}

\def\z{{\zeta}}
\def\ap{\alpha'}
\def\zbar{{\bar z}}

\def\cf{{\it cf.\ }}
\def\ie{{\it i.e.\ }}
\def\eg{{\it e.g.\ }}
\def\eqq{{\it Eq. \ }}

\def\bet{\beta}

\newif\ifnref

\def\doubref#1#2{\refs{{#1},{#2} }}

\nreffalse

\input epsf

\def\figin{\epsfcheck\figin}\def\figins{\epsfcheck\figins}
\def\epsfcheck{\ifx\epsfbox\UnDeFiNeD
\message{(NO epsf.tex, FIGURES WILL BE IGNORED)}
\gdef\figin##1{\vskip2in}\gdef\figins##1{\hskip.5in}
\else\message{(FIGURES WILL BE INCLUDED)}%
\gdef\figin##1{##1}\gdef\figins##1{##1}\fi}
\def\DefWarn#1{}
\def\figinsert{\goodbreak\midinsert}  
\def\ifig#1#2#3{\DefWarn#1\xdef#1{Fig.~\the\figno}
\writedef{#1\leftbracket fig.\noexpand~\the\figno}%
\figinsert\figin{\centerline{#3}}\medskip\centerline{\vbox{\baselineskip12pt
\advance\hsize by -1truein\noindent\footnotefont\centerline{{\bf
Fig.~\the\figno}\ \sl #2}}}
\bigskip\endinsert\global\advance\figno by1}

\def\iifig#1#2#3#4{\DefWarn#1\xdef#1{Fig.~\the\figno}
\writedef{#1\leftbracket fig.\noexpand~\the\figno}%
\figinsert\figin{\centerline{#4}}\medskip\centerline{\vbox{\baselineskip12pt
\advance\hsize by -1truein\noindent\footnotefont\centerline{{\bf
Fig.~\the\figno}\ \ \sl #2}}}\smallskip\centerline{\vbox{\baselineskip12pt
\advance\hsize by -1truein\noindent\footnotefont\centerline{\ \ \ \sl #3}}}
\bigskip\endinsert\global\advance\figno by1}


\def\appA{A}
\def\appB{B}

\def\tilde{\widetilde}

\def\h {{1\over 2}}
\def\hb{{\bar h}}
\def\ov {\overline}
\def\o {\over}
\def\fc#1#2{{#1 \o #2}}

\def\IP{{\bf P}}\def\IC{{\bf C}}\def\IR{ {\bf R}}


\def\br{\hfill\break}

\def\lf {\left}
\def\ri {\right}
\def\ra {\rightarrow}

\def\re {{\rm Re}}

\def\p {\partial}

 \def\Oc {{\cal O}}
\def\Lc {{\cal L}}

\def\Hc{{\cal H}}


\def\shuffle{{\hskip0.10cm \vrule height 0pt width 8pt depth 0.75pt  \hskip-0.3cm\ss{\rm III}\hskip0.05cm}}
\def\sv{{\rm sv}}
\def\SVM{{\zeta^m_{\rm sv}}}
\def\SV{{\zeta_{\rm sv}}}


\lref\GreenXX{
  M.B.~Green and J.H.~Schwarz,
``Supersymmetrical Dual String Theory. 2. Vertices and Trees,''
Nucl.\ Phys.\ B {\bf 198}, 252 (1982).
}
\lref\GrossRR{
  D.J.~Gross, J.A.~Harvey, E.J.~Martinec and R.~Rohm,
``Heterotic String Theory. 2. The Interacting Heterotic String,''
Nucl.\ Phys.\ B {\bf 267}, 75 (1986).
}

\lref\BrownGIA{
  F.~Brown,
``Single-valued Motivic Periods and Multiple Zeta Values,''
SIGMA {\bf 2}, e25 (2014).
[arXiv:1309.5309 [math.NT]].
}

\lref\StiebergerWEA{
  S.~Stieberger,
``Closed superstring amplitudes, single-valued multiple zeta values and the Deligne associator,''
J.\ Phys.\ A {\bf 47}, 155401 (2014).
[arXiv:1310.3259 [hep-th]].
}

\lref\StromingerZOO{
  A.~Strominger,
  ``Lectures on the Infrared Structure of Gravity and Gauge Theory,''
[arXiv:1703.05448 [hep-th]].
}

\lref\PasterskiYLZ{
  S.~Pasterski, S.H.~Shao and A.~Strominger,
 ``Gluon Amplitudes as 2d Conformal Correlators,''
Phys.\ Rev.\ D {\bf 96}, no. 8, 085006 (2017).
[arXiv:1706.03917 [hep-th]].
}

\lref\LysovCSA{
  V.~Lysov, S.~Pasterski and A.~Strominger,
  ``Low's Subleading Soft Theorem as a Symmetry of QED,''
Phys.\ Rev.\ Lett.\  {\bf 113}, no. 11, 111601 (2014).
[arXiv:1407.3814 [hep-th]].
}
\lref\Kolbig{
K.S. K\"olbig,
``Nielsen's Generalized Polylogarithms,''
SIAM J. Math. Anal. 17, 1232-1258, 1986.}

\lref\Kolbigi{
K.S. K\"olbig,
``Closed expressions for $\int_0^1t^{-1}\ln^{n-1}t\ \ln^p(1-t)\ dt$,''
Mathematics of Computation 39 (160) (1982)
647Ð654.}

\lref\KolbigZZA{
  K.S.~K\"olbig, J.A.~Mignoco and E.~Remiddi,
``On Nielsen's Generalized Polylogarithms And Their Numerical Calculation,''
CERN-DD-CO-69-5.
}
\lref\SchreiberJSR{
  A.~Schreiber, A.~Volovich and M.~Zlotnikov,
  ``Tree-level gluon amplitudes on the celestial sphere,''
Phys.\ Lett.\ B {\bf 781}, 349 (2018).
[arXiv:1711.08435 [hep-th]].
}
\lref\Adamchik{
V. Adamchik,
``On Stirling numbers and Euler sums,''
 J. Comput. Appl. Math. {\bf 79} (1997) 119--130.
}
\lref\PasterskiQVG{
  S.~Pasterski, S.H.~Shao and A.~Strominger,
  ``Flat Space Amplitudes and Conformal Symmetry of the Celestial Sphere,''
Phys.\ Rev.\ D {\bf 96}, no. 6, 065026 (2017).
[arXiv:1701.00049 [hep-th]].
}

\lref\Sommerfeld{
A. Sommerfeld, ``Partial Differential Equations in Physics,'' Academic Press New York (1949) pp 279-289;\br
G.N. Watson, ``The diffraction of electric waves by the earth,'' Proc. R. Soc. Lond. A {\bf 95} (1918)  83-99.}
\lref\WittenPRA{
  E.~Witten,
  ``The Feynman $i \epsilon$ in String Theory,''
JHEP {\bf 1504}, 055 (2015).
[arXiv:1307.5124 [hep-th]].}

\lref\StiebergerHBA{
  S.~Stieberger and T.R.~Taylor,
``Closed String Amplitudes as Single-Valued Open String Amplitudes,''
Nucl.\ Phys.\ B {\bf 881}, 269 (2014).
[arXiv:1401.1218 [hep-th]].
}
\lref\TaylorSPH{
  T.R.~Taylor,
  ``A Course in Amplitudes,''
Phys.\ Rept.\  {\bf 691}, 1 (2017).
[arXiv:1703.05670 [hep-th]].
}

\lref\GrossAR{
  D.J.~Gross and P.F.~Mende,
``String Theory Beyond the Planck Scale,''
Nucl.\ Phys.\ B {\bf 303}, 407 (1988);
``The High-Energy Behavior of String Scattering Amplitudes,''
Phys.\ Lett.\ B {\bf 197}, 129 (1987).
}

\lref\Gonch{
A.B. Goncharov, ``Multiple polylogarithms, cyclotomy and modular complexes,'' Math.
Research Letters, 5 (1998), 497--516 [arXiv:1105.2076];
``Multiple polylogarithms and mixed Tate motives,'' 2001
[math/0103059v4].}

\lref\StiebergerHZA{
  S.~Stieberger and T.R.~Taylor,
  ``Superstring Amplitudes as a Mellin Transform of Supergravity,''
Nucl.\ Phys.\ B {\bf 873}, 65 (2013).
[arXiv:1303.1532 [hep-th]].
}
\lref\BernQJ{
  Z.~Bern, J.J.M.~Carrasco and H.~Johansson,
  ``New Relations for Gauge-Theory Amplitudes,''
Phys.\ Rev.\ D {\bf 78}, 085011 (2008).
[arXiv:0805.3993 [hep-ph]].
}
\lref\StiebergerNHA{
  S.~Stieberger and T.R.~Taylor,
  ``Superstring/Supergravity Mellin Correspondence in Grassmannian Formulation,''
Phys.\ Lett.\ B {\bf 725}, 180 (2013).
[arXiv:1306.1844 [hep-th]].
}

\lref\BrownUGM{
  F.C.S.~Brown,
``Polylogarithmes multiples uniformes en une variable,''
Compt.\ Rend.\ Math.\  {\bf 338}, no. 7, 527 (2004)..
}

\lref\GDR{
  C.~Duhr, H.~Gangl and J.R.~Rhodes,
``From polygons and symbols to polylogarithmic functions,''
JHEP {\bf 1210}, 075 (2012).
[arXiv:1110.0458 [math-ph]].
}

\lref\Gonch{
A.B. Goncharov, ``Multiple polylogarithms, cyclotomy and modular complexes,'' Math.
Research Letters, 5 (1998), 497--516 [arXiv:1105.2076];
``Multiple polylogarithms and mixed Tate motives,'' 2001
[math/0103059v4].}

\lref\PasterskiKQT{
  S.~Pasterski and S.H.~Shao,
  ``Conformal basis for flat space amplitudes,''
Phys.\ Rev.\ D {\bf 96}, no. 6, 065022 (2017).
[arXiv:1705.01027 [hep-th]].
}

\lref\SchnetzHQA{
  O.~Schnetz,
``Graphical functions and single-valued multiple polylogarithms,''
Commun.\ Num.\ Theor.\ Phys.\  {\bf 08}, 589 (2014).
[arXiv:1302.6445 [math.NT]].
}

\lref\FanUQY{
  W.~Fan, A.~Fotopoulos, S.~Stieberger and T.R.~Taylor,
``SV-map between Type I and Heterotic Sigma Models,''
Nucl.\ Phys.\ B {\bf 930}, 195 (2018).
[arXiv:1711.05821 [hep-th]].
}
\lref\StiebergerQJA{
  S.~Stieberger and T.R.~Taylor,
  ``Graviton Amplitudes from Collinear Limits of Gauge Amplitudes,''
Phys.\ Lett.\ B {\bf 744}, 160 (2015).
[arXiv:1502.00655 [hep-th]].
}
\lref\StiebergerCEA{
  S.~Stieberger and T.R.~Taylor,
  ``Graviton as a Pair of Collinear Gauge Bosons,''
Phys.\ Lett.\ B {\bf 739}, 457 (2014).
[arXiv:1409.4771 [hep-th]].
}

\lref\StiebergerVYA{
  S.~Stieberger and T.R.~Taylor,
``Disk Scattering of Open and Closed Strings (I),''
Nucl.\ Phys.\ B {\bf 903}, 104 (2016).
[arXiv:1510.01774 [hep-th]].
}

\lref\VenezianoYB{
  G.~Veneziano,
  ``Construction of a crossing - symmetric, Regge behaved amplitude for linearly rising trajectories,''
Nuovo Cim.\ A {\bf 57}, 190 (1968).
}

\lref\deBoerVF{
  J.~de Boer and S.N.~Solodukhin,
  ``A Holographic reduction of Minkowski space-time,''
Nucl.\ Phys.\ B {\bf 665}, 545 (2003).
[hep-th/0303006].
}
\lref\CheungIUB{
  C.~Cheung, A.~de la Fuente and R.~Sundrum,
  ``4D scattering amplitudes and asymptotic symmetries from 2D CFT,''
JHEP {\bf 1701}, 112 (2017).
[arXiv:1609.00732 [hep-th]].
}

\lref\StiebergerLNG{
  S.~Stieberger and T.R.~Taylor,
``New relations for Einstein-Yang-Mills amplitudes,''
Nucl.\ Phys.\ B {\bf 913}, 151 (2016).
[arXiv:1606.09616 [hep-th]].
}
\lref\KawaiXQ{
  H.~Kawai, D.C.~Lewellen and S.H.H.~Tye,
  ``A Relation Between Tree Amplitudes of Closed and Open Strings,''
Nucl.\ Phys.\ B {\bf 269}, 1 (1986).
}
\lref\SchlottererCXA{
  O.~Schlotterer,
``Amplitude relations in heterotic string theory and Einstein-Yang-Mills,''
JHEP {\bf 1611}, 074 (2016).
[arXiv:1608.00130 [hep-th]].
}
\lref\GRAV{
  S.~Stieberger,
 ``Constraints on Tree-Level Higher Order Gravitational Couplings in Superstring Theory,''
Phys.\ Rev.\ Lett.\  {\bf 106}, 111601 (2011)
[arXiv:0910.0180 [hep-th]].
}

\lref\SS{
  O.~Schlotterer and S.~Stieberger,
``Motivic Multiple Zeta Values and Superstring Amplitudes,''
J.\ Phys.\ A {\bf 46}, 475401 (2013).
[arXiv:1205.1516 [hep-th]].
}

\Title{\vbox{\rightline{MPP--2018--136}
}}
{\vbox{\centerline{Strings on Celestial Sphere}
}}
\medskip
\centerline{Stephan Stieberger$^{a,b}$ ~,~ Tomasz R. Taylor$^{c,d}$}
\bigskip
\centerline{\it $^a$ Max--Planck--Institut f\"ur Physik}
\centerline{\it Werner--Heisenberg--Institut, 80805 M\"unchen, Germany}
\medskip
\centerline{\it $^b$ Kavli Institute for Theoretical Physics}
\centerline{\it University of California, Santa Barbara, CA 93106, USA}
\medskip
\centerline{\it $^c$ Institute of Theoretical Physics, Faculty of Physics}
\centerline{\it University of Warsaw, Poland}
\medskip
\centerline{\it  $^d$ Department of Physics}
\centerline{\it  Northeastern University, Boston, MA 02115, USA}
\vskip15pt

\vskip15pt

\medskip
\bigskip\bigskip\bigskip
\centerline{\bf Abstract}
\vskip .2in
\noindent
We transform superstring scattering amplitudes into the correlation functions of primary conformal fields on two-dimensional celestial sphere. The points on celestial sphere are associated to the asymptotic directions of (light--like) momenta of external particles, with the Lorentz group realized as the $SL(2,\IC)$ conformal symmetry of the sphere. The energies are dualized through Mellin transforms into the parameters that determine dimensions of the primaries. We focus on four--point amplitudes involving gauge bosons and gravitons in type I open superstring theory and in closed heterotic superstring theory at the tree--level.

\noindent

\Date{}
\noindent
\goodbreak
\listtoc
\writetoc
\break
\newsec{Introduction}
In modern-day particle colliders, accelerators produce beams of incident particles with specific energies and momenta, described to a reasonable accuracy by the packets of plane waves with a narrow spread of four-momentum. Similarly, the detectors are designed to measure four-momenta of the scattered particles. Hence it is not surprising that almost all studies of the scattering amplitudes are focussed on the transitions between four-momentum eigenstates (planar wave-functions). For example, the Feynman rules are usually formulated in such a momentum representation.

Much of the recent progress in computing the scattering amplitudes is due to the applications of spinor-helicity techniques. For a review, notations and conventions, see \TaylorSPH. The amplitudes describing massless particles are most succinctly expressed in terms of momentum spinors which transform under Lorentz transformations in the defining representations of $SL(2,\IC)$. Some time ago, when investigating relations between field-theoretical and string amplitudes, we defined complex projective coordinates $z\equiv \sigma_1/\sigma_2$, the ratios of momentum spinor components $\sigma_{1}$ and $\sigma_{2}$, and mapped these kinematic variables into the positions of vertex operators on the Riemann sphere describing string world-sheet \refs{\StiebergerHZA,\StiebergerNHA}. The Lorentz symmetry under
\eqn\conf{z\to{az+b\over cz+d}\qquad (ad-bc=1)\ ,}
was mapped into conformal symmetry group of the spherical world-sheet. This mirrored the observation made long ago by Penrose that the snapshots of night sky -- the so-called celestial sphere, as taken by different observers, are related by such conformal transformations.

More recently, Strominger \StromingerZOO\ and collaborators applied a similar construction to map the scattering amplitudes  from the momentum space  to celestial sphere in Refs.\ \refs{\PasterskiQVG,\PasterskiYLZ}. In Minkowski spacetime parameterized by Bondi coordinates $(u,r,z,\bar z)$,  $z$ and $\zbar$ describe celestial sphere. On the other hand, in terms of the projective coordinates mentioned in the previous paragraph, any light-like momentum can be written as
\eqn\mom{p^\mu=\omega q^\mu\ ,\quad {\rm with}~~ q^\mu={1\over 2}\ (1+|z|^2, z+\zbar,-i(z-\zbar),1-|z|^2)\ ,}
where $\omega$ is the light-cone energy scale which transforms as \eqn\ener{\omega\to(cz+d)\ (\bar c\zbar+\bar d)\ \omega}
under conformal transformations \conf.
After expressing all kinematic variables in terms of $\omega$, $z$ and $\zbar$, the standard transition amplitudes between momentum eigenstates become functions of celestial coordinates and energies. Actually, the amplitudes can be streamlined into familiar 2D CFT correlators by considering the scattering of so-called conformal wave packets \PasterskiKQT
\eqn\wp{\varphi_{\Delta}^\pm(x^\mu;z,\zbar)=\int_0^\infty\omega^{\Delta-1}e^{\pm i\omega q\cdot x-\epsilon\omega}={(\mp i)^\Delta\Gamma(\Delta)\over [-x\cdot q(z,\zbar)\mp i\epsilon]^\Delta}\ ,}
which are Mellin transforms of the usual plane waves. These packets are described by massless scalar conformal primary wave functions of dimension $\Delta$, the variable dual to the energy in the Mellin sense, and can be generalized to higher spin \PasterskiKQT.  By using such Mellin transforms, ``old-fashioned'' gauge and gravitational amplitudes can be converted into conformal correlators of primary fields on celestial sphere, labeled by their conformal spin and dimensions.

There are several interesting aspects of this proposal. Perhaps most remarkably, understanding the nature of 2D CFT on celestial sphere would enable a holographic description of flat spacetime \refs{\deBoerVF,\CheungIUB}. Unfortunately, as pointed out by Strominger, it is not a ``garden variety" of CFT, although it has some intriguing properties. For example, soft photons correspond to $\Delta=1$ current insertions on celestial sphere, and the related soft theorems can be interpreted as Ward identities associated to certain asymptotic symmetries \refs{\StromingerZOO,\LysovCSA}. We are interested, however, in  another aspect.

Together with the progress in computing the scattering amplitudes in perturbative gauge theories, Einstein's gravity, Einstein-Yang-Mills (EYM) theory and string theory, it became clear that gravitational interactions are closely related to gauge interactions, at least at lowest orders of perturbation theory. The archetypal Kawai-Llewellen-Tye (KLT) relations relate the amplitudes with external gravitons to products of pure gauge amplitudes \KawaiXQ. The color--kinematic duality reveals some intriguing kinematic gauge structures hidden in gravitational amplitudes \BernQJ. Finally, the most recent collinear relations \refs{\StiebergerCEA,\StiebergerQJA,\StiebergerVYA} allow substituting gravitons with pairs of collinear gauge bosons in practically all EYM amplitudes. All this indicates to yet unknown, profound connections between gravity and gauge interactions. Can 2D CFT on celestial sphere offer some new insight into these connections?

There are some good reasons to address the above question in the context of string theory. Indeed, the gauge-gravity connection appears to be a consequence of the observation that closed strings look like two open strings connected at the ends. The most important reason however is that every order in the perturbative expansion of gravity violates the unitarity bounds by growing powers of energy. As we will see later, this uncontrollable growth at large energies poses an obstacle for transforming gravitational amplitudes to celestial sphere. This problem does not appear in string amplitudes which are renowned for their super-soft ultraviolet behavior \refs{\VenezianoYB,\GrossAR}. Furthermore,
returning to the connection of kinematic variables to the string world-sheet mentioned at the beginning, we want to know if there are any connections between a relatively simple CFT on the world-sheet and rather intricate CFT on celestial sphere. To that end, we will discuss Mellin transforms of full-fledged superstring amplitudes. We will start from the amplitudes with three external particles, for which there is no difference between QFT and string theory.
\newsec{Preliminaries and three-particle amplitudes}
In the first step towards celestial life, the amplitudes are expressed in terms of complex coordinates and light-cone energies. The relevant $SL(2,\IC)$ conformal transformation properties are given in Eqs.\conf\ and \ener, respectively. We will be classifying conformal primary fields according to their conformal weights $(h,\hb)$ or their dimensions $\Delta=h+\hb$ and spins $J=h-\hb$.
Note that energies transform in \ener\ as weight $(1/2,1/2)$ primaries. The amplitudes, written in the helicity basis, depend on the spinor products \TaylorSPH\ \eqn\spin{\langle ij\rangle=\sqrt{\omega_i\omega_i}\ z_{ij}~,\qquad
[ij]=-\sqrt{\omega_i\omega_j}\ \zbar_{ij}\qquad (z_{ij}\equiv z_i-z_j,~ \zbar_{ij}\equiv \zbar_i-\zbar_j)}
and the usual scalar products: \eqn\mom{s_{ij}\equiv 2p_ip_j=\langle ij\rangle[ji]=\omega_i\omega_j z_{ij}\zbar_{ij}\ .}
The angle products $\langle\cdots\rangle$ have weights $(-1/4,1/4)$ while the square products $[\cdots]$ have weights $(1/4,-1/4)$. These weights allow identifying four-dimensional helicity with 2D conformal spin.

We will start from two-particle collisions in which two incident particles, with momenta $p_1$ and $p_2$, scatter into $N{-}2$ particles in the final state. They are described by the amplitudes of the general form
\eqn\amp{{\cal A}=i(2\pi)^4\ \delta^{(4)}\lf(p_1+p_2-\sum_{k=3}^Np_k\ri)\  {\cal M}\ ,}
where ${\cal M}$ are the so-called invariant matrix elements that can be computed by using Feynman rules or some other techniques\foot{From now on we will skip the factor $i(2\pi)^4$.}. They depend on all quantum numbers, including internal gauge charges, and may contain some group-dependent (a.k.a.\ color) factors. In such cases, {\it i.e}.\ in the presence of gauge bosons,  we will be considering the so-called ``partial'' amplitudes associated to the canonical trace factor, that is purely kinematic functions ``stripped'' of color factors.

All conformal primary wave functions \wp\ and their higher spin partners solve 4D wave equations, but in order to form complete, normalizable sets,  their dimensions must be restricted to the principal continuous series with $\Delta=1+i\lambda$, $\lambda\in\IR$ \PasterskiKQT. Thus Mellin transforming a given amplitude from the momentum basis to the conformal basis amounts to evaluating:
\eqn\melt{\tilde{\cal A}_{\{\lambda_n\}}(z_n,\zbar_n)=\bigg(\prod_{n=1}^{N}
\int_0^\infty\omega_n^{i\lambda_n}d\omega_n\bigg)\ \delta^{(4)}(\omega_1 q_1+\omega_2 q_2-\sum_{k=3}^N\omega_k q_k)\  {\cal M}(\omega_n,z_n,\zbar_n)\ .}

{}In the case of three external particles, the amplitudes vanish because the constraints of momentum conservation, as enforced by the delta function inside \melt, force all kinematic invariants to be zero. These constraints can be relaxed by changing the metric signature from $(+--\, -)$ to $(++-\, -)$. This allows treating $z$ and $\zbar$ as two {\it independent\/} real variables. Then two classes of non-trivial kinematic solutions are allowed: all $z_{ij}=0$ with all $\zbar_{ij}\neq 0$ or all $\zbar_{ij}=0$ with all $z_{ij}\neq 0$. In the case of  amplitudes involving three gauge bosons, the first one is appropriate for  ``mostly minus'' helicity configurations while the second one is good for the ``mostly plus'' amplitudes. We will focus on the latter ones. Assuming all $z_{ij}\neq 0$, the momentum-conserving delta function can be written as
\eqn\dmelt{\delta^{(4)}(\omega_1 q_1+\omega_2 q_2-\omega_3 q_3) ={4\over \omega_3^2}\ {1\over z_{23}z_{31}}\ \delta(\omega_1-{z_{32}\over z_{12}}\omega_3)\ \delta(\omega_2-{z_{31}\over z_{21}}\omega_3)\ \delta(\zbar_{13})\ \delta(\zbar_{23})\ ,}
with the additional constraint that the variables must be ordered in one of two possible ways:
$z_1< z_3<z_2$ or $z_2< z_3<z_1$, to ensure that all energies are positive.

The mostly-plus three-gluon amplitude is given by\foot{We are using here a self-explanatory notation and skip the coupling constant factors. In case of any doubt, the reader should consult \TaylorSPH.}:
\eqn\glu{{\cal M}(-,-,+)={\langle 12\rangle^3\over \langle 13\rangle\langle 23\rangle}=
{\omega_1\omega_2\over\omega_3}{z_{12}^3\over z_{13}z_{23}}\ .}
The corresponding celestial amplitude is:
\eqn\cglu{\tilde{\cal A}(-,-,+)=4\,{z_{21}^{1-i(\lambda_1+\lambda_2)}z_{23}^{i\lambda_1-1}z_{31}^{i\lambda_2-1} }\,
\delta(\zbar_{13})\delta(\zbar_{23})\int_0^\infty \omega_3^{i(\lambda_1+\lambda_2+\lambda_3)-1}d\omega_3\ .}
This amplitude has conformal transformation properties of a three-point correlation function of primary conformal fields with weights\foot{Actually, due to the delta functions, $\hb$-weights are not uniquely determined. The only constraint is $\sum_{n=1}^3\hb_n=2+{i\over 2}\sum_{n=1}^3\lambda_n$.}
\eqn\weight{\eqalign{&h_1={i\over 2}\lambda_1,\qquad\qquad \hb_1=1+{i\over 2}\lambda_1,\cr
&h_2={i\over 2}\lambda_2,\qquad\qquad \hb_2=1+{i\over 2}\lambda_2,\cr
&h_3=1+{i\over 2}\lambda_3,\qquad ~\hb_3={i\over 2}\lambda_3,}}
in agreement with $\Delta_n=1+i\lambda_n$, $J_1=J_2=-1$ and $J_3=+1$.
Note that the energy integral remaining on the r.h.s\ of \cglu\ is logarithmically divergent in the infra-red and in ultra-violet. Since any cutoff would violate $SL(2,\IC)$ symmetry, there is no other choice than to interpret it in the sense of a distribution \PasterskiYLZ:
\eqn\distr{\int_0^\infty \omega_3^{i(\lambda_1+\lambda_2+\lambda_3)-1}d\omega_3=2\pi\ \delta(
\lambda_1+\lambda_2+\lambda_3)\ .}
We do not discuss here other helicity configurations because they can be analysed in a similar way.

The mostly-plus three-graviton amplitude is given by the square of three-gluon amplitude\foot{Here again, we skip the (gravitational) coupling constant factors.}:
\eqn\gru{{\cal M}(--,--,++)={\langle 12\rangle^6\over \langle 13\rangle^2\langle 23\rangle^2}=
{\omega_1^2\omega_2^2\over\omega_3^2}{z_{12}^6\over z_{13}^2z_{23}^2}\ .}
The corresponding celestial amplitude is:
\eqn\cgru{\tilde{\cal A}(--,--,++)=4\,{z_{21}^{2-i(\lambda_1+\lambda_2)}z_{23}^{i\lambda_1-1}z_{31}^{i\lambda_2-1} }\,
\delta(\zbar_{13})\ \delta(\zbar_{23})\int_0^\infty \omega_3^{i(\lambda_1+\lambda_2+\lambda_3)}d\omega_3\ .}
It has conformal transformation properties of a three-point correlation function of primary conformal fields with weights\foot{Actually, due to the delta functions, $\hb$-weights are not uniquely determined. The only constraint is $\sum_{n=1}^3\hb_n=2+{i\over 2}\sum_{n=1}^3\lambda_n$.}
\eqn\gweight{\eqalign{&h_1=-{1\over 2}+{i\over 2}\lambda_1,\qquad\qquad \hb_1={3\over 2}+{i\over 2}\lambda_1,\cr
&h_2=-{1\over 2}+{i\over 2}\lambda_2,\qquad\qquad \hb_2={3\over 2}+{i\over 2}\lambda_2,\cr
&h_3={3\over 2}+{i\over 2}\lambda_3,\qquad\qquad ~~\,\hb_3=-{1\over 2}+{i\over 2}\lambda_3\ ,}}
in agreement with $\Delta_n=1+i\lambda_n$, $J_1=J_2=-2$ and $J_3=+2$. The main difference, however, between the gravitational and gauge amplitudes is the energy integral, which  in the gravitational case \cgru\ is linearly divergent in the ultraviolet. The degree of this divergence will grow with the number of external gravitons, reflecting the violation of unitarity bounds at each order of perturbative Einstein's gravity. One needs an ultraviolet completion of the theory in order to make sense out of the gravitational amplitude \gru.  In the next section, we will see that string theory does indeed provide such a completion.

There is one more mostly-plus amplitude which is useful for studying the gauge-gravity connection. It is the EYM amplitude involving one graviton and two gauge bosons
\eqn\gyu{{\cal M}(--,-,+)={\langle 12\rangle^4\over \langle 23\rangle^2}=
{\omega_1^2\omega_2\over\omega_3}{z_{12}^4\over z_{23}^2}\ .}
The corresponding celestial amplitude is
\eqn\cgyu{\tilde{\cal A}(--,-,+)=4\,{z_{21}^{1-i(\lambda_1+\lambda_2)}\ z_{23}^{i\lambda_1-1}\ z_{31}^{i\lambda_2} }\,
\delta(\zbar_{13})\ \delta(\zbar_{23})\ \int_0^\infty \omega_3^{i(\lambda_1+\lambda_2+\lambda_3)}d\omega_3\ ,}
which is linearly divergent again. It is easy to check that also in this case, the weights agree with the dimensions and spins of gauge and graviton primaries.

In superstring theory, at the energies much smaller than the characteristic string energy scale (determined by the `universal Regge slope'' parameter $\alpha'$),
gravitons and gauge boson interact exactly the same way as in EYM theory. String corrections due to massive string excitations appear at higher energies and are often discussed in terms of the expansion in powers of $\alpha'$, which has dimension of length square. These corrections are absent, however, at the level of three-point amplitudes, for purely kinematic reasons. In the next section, we will see how string effects appear in four-particle celestial amplitudes.

\newsec{Four--gluon amplitudes in open superstring theory}
At the perturbative level, two distinct superstring theories include massless gauge bosons: type I open superstrings and heterotic superstrings. The latter incorporates gravitons in the massless spectrum and is suitable for studying mixed gauge-gravitational amplitudes. Virtual gravitons and massive neutral closed string excitations can propagate also inside pure gauge amplitudes, therefore the heterotic theory gives rise to a richer variety of multi-trace color structures. Here, we focus on the amplitudes with single-trace color factors and the corresponding partial amplitudes. Open and heterotic single trace amplitudes are different, but they both reproduce Yang Mills amplitudes in the $\alpha'=0$ limit. Furthermore, they are related by the mathematical operation called ``single-value'' (sv) projection \refs{\BrownGIA,\StiebergerWEA}:
the $\alpha'$ expansion series of heterotic amplitudes can be obtained by acting with sv on the open ones \refs{\StiebergerHBA,\FanUQY}. The effect of single--value projection is to map
the zeta function coefficients onto a subspace thereof, for example sv$[\zeta(2)]=0$, sv$[\zeta(3)]=2\zeta(3)$, etc. The kinematic dependence of $\alpha'$ expansion coefficients remains untouched by the map sv.
{}For the purpose of our discussion, we will be considering  open and heterotic cases separately, in each case applying \melt\ to transform the amplitudes.

In the case of four particles, the momentum-conserving delta function inside \melt\ can be rewritten as
\eqn\dmeltf{\eqalign{\delta^{(4)}(\omega_1 q_1+\omega_2 q_2&-\omega_3 q_3-\omega_4 q_4)
={4\over \omega_4|z_{14}|^2|z_{23}|^2}\cr & \times\, \delta\lf(\omega_1-{z_{24}\zbar_{34}\over z_{12}\zbar_{13}}\omega_4\ri)\ \delta\lf(\omega_2-{z_{14}\zbar_{34}\over z_{12}\zbar_{32}}\omega_4\ri)\ \delta\lf(\omega_3+{z_{24}\zbar_{14}\over z_{23}\zbar_{13}}\omega_4\ri)\,
\delta(r-\bar r)\ ,}}
where $r$ is the conformal invariant cross ratio:
\eqn\rat{r={z_{12}z_{34}\over z_{23}z_{41}}\ .}
The physical meaning of this parameter can be understood by computing the ratio of Mandelstam's  variables $s=s_{12}=(p_1+p_2)^2$ and $u=-s_{23}=(p_2-p_3)^2$
\eqn\mand{{s_{23}\over s_{12}}={1\over r}=-{u\over s}=\sin^2\!\bigg({\theta\over 2}\bigg)\ ,}
where $\theta$ is the scattering angle in the center of mass frame. From the above relation it is clear that the cross-ratios are restricted to $r>1$. Indeed, $r=\bar r$, as required by the delta function in \dmeltf, together with $r>1$, ensure that all energies, as determined by the delta functions remaining in \dmeltf, are real and positive. In the physical domain, $s>0$ and $u=-s/r<0$.

All four--particle amplitudes belong to the class of maximally helicity violating (MHV); as in the previous section, we will be considering mostly plus MHV configurations only. The well-known four-gluon Yang-Mills amplitude is
\eqn\ymhv{{\cal M}(-,-,+,+)={\langle 12\rangle^3\over \langle 23\rangle\langle 34\rangle\langle 41\rangle}={\omega_1\omega_2\over\omega_3\omega_4}{z_{12}^3\over z_{23}z_{34}z_{41}}=r\,{z_{12}
\zbar_{34}\over \zbar_{12}z_{34}}\ ,}
where in the last step, we used the constraints of \dmeltf. The corresponding celestial amplitude is
\eqn\cgluf{\eqalign{ \tilde{\cal A}(-,-,+,+) & =4\ \delta(r-\bar r)\,\bigg({\zbar_{34}\over z_{12}}\bigg)^{i\lambda_1}\bigg({z_{34}\over \zbar_{12}}\bigg)^{i\lambda_2}\ \bigg({z_{24}\over \zbar_{13}}\bigg)^{i(\lambda_1+\lambda_3)}\bigg({\zbar_{14}\over z_{23}}\bigg)^{i(\lambda_2+\lambda_3)}\cr &\times\theta(r-1)\,{r^3\over\zbar_{12}^2\, z_{34}^2}\, J_0\ ,}}
where the step function $\theta(r-1)$ enforces the kinematic constraint $r>1$ and the energy integral is:
\eqn\jzero{J_0=\int_0^\infty \omega_4^{i(\lambda_1+\lambda_2+\lambda_3+\lambda_4)-1}d\omega_4\ .}
It is easy to see that the conformal weights agree with $\Delta_n=1+i\lambda_n$, $J_1=J_2=-1$ and $J_3=J_4=+1$. As in the three-particles case, the energy integral yields:
\eqn\jzeroq{J_0=2\pi\ \delta\bigg(\sum_{n=1}^4\lambda_n\bigg)\ .}
At this point, one can cast the amplitude in a form appropriate to a four-point CFT correlator
\eqn\cglug{\eqalign{ \tilde{\cal A}(-,-,+,+) &=8\pi \ \delta(r-\bar r)\,\delta\bigg(\sum_{n=1}^4\lambda_n\bigg)
\cr
&\times\Bigg(\prod_{i<j}^4 z_{ij}^{{h\over 3}-h_i-h_j}
\zbar_{ij}^{{\bar h\over 3}-\bar h_i-\bar h_j}\Bigg)\, r^{5\over 3}\,(r-1)^{2\over 3}\, \theta(r-1)\ ,}}
where $h=\sum_{n=1}^4h_n$ and $\bar h=\sum_{n=1}^4\bar h_n$. Up to some numerical factors, the above result is in agreement\foot{Yang-Mills amplitudes with five and more external gluons have been recently discussed in~\SchreiberJSR.}  with \PasterskiYLZ.

The type I open superstring amplitude is related to the Yang-Mills amplitude by a simple rescaling
\eqn\smhv{{\cal M}_I(-,-,+,+)={\cal M}(-,-,+,+)\ F_I(s,u)\ ,}
with the string ``formfactor" \GreenXX
\eqn\formf{F_I(s,u)=-\ap s_{12}B(-\ap s_{12},1+\ap s_{23})=-sB(-s,1-u)={\Gamma (1-s)\Gamma(1-u)\over\Gamma(1-s-u)}\ ,}
where we rescaled Mandelstam's variables by the the string scale:
$s\equiv\ap s_{12}$ and  $u\equiv -\ap s_{23}=-s/r$. Since $s>0$ and $u<0$, the poles due to massive string excitations appear in the $s$-channel only, at $s$ integer and positive.
Before transforming the amplitude to celestial sphere, we want to exhibit the well-known super-soft high energy behavior of the formfactor, at $s\to \infty$ and fixed angle, {\it i.e}.\ fixed $r>1$.
It is convenient to define $a=1/r\in (0,1)$, so that $u=-as$. Then
\eqn\formfas{F_I(s,u)=as\ {\sin[\pi(1-a)s]\over\sin(\pi s)}\ {\Gamma (as)\ \Gamma[(1-a)s]\over\Gamma(s)}}
and the asymptotic behavior can be determined by using Stirling's formula
\eqn\formfat{F_I(s,u)\sim\sqrt{2\pi as\over(1-a)}\ {\sin[\pi(1-a)s]\over\sin(\pi s)}
\  a^{as}(1-a)^{(1-a)s}\ ,}
which is exponentially suppressed at $s\to\infty$, except at the singular points.

The celestial string amplitude corresponding to \smhv\ is given by the same expression as the Yang-Mills amplitude \cgluf, but now with $J_0$ replaced by a non-trivial energy integral:
\eqn\jzeri{J_I=\int_0^\infty \omega_4^{i(\lambda_1+\lambda_2+\lambda_3+\lambda_4)-1}F_I(s,u)\,d\omega_4\ .}
Recall that the formfactor $F_I$ is given by \formf, with:
\eqn\mands{s=\ap(r-1)\, {|z_{14}|^2|z_{34}|^2\over |z_{13}|^2}\,\omega_4^2\ ,\qquad u=-{s\over r}\ .}
The basic difference between Yang-Mills \jzero\ and string \jzeri\ energy integrals is  in the ultra-violet regime at $\omega_4\to\infty$, where the exponential suppression \formfat\ of the string formfactor makes it square-integrable. In addition, the integration runs over massive string poles, although it is not a problem because these singularities can be handled by using the $i\epsilon$ prescription \WittenPRA. Note that there is no difference in the infrared because $F(s,u)\to 1$ as $\omega_4\to 0$. It is convenient to change the integration variables and express the amplitude in terms of the integral:
\eqn\ibeta{
I(r,\beta):={1\over 2}\int_0^\infty  w^{-\beta-1}F_I(rw,-w)\,dw~,\qquad\beta:= -{i\over 2}
\sum_{n=1}^4\lambda_n\ .}
After some algebra, we obtain:
\eqn\cglus{\eqalign{ \tilde{\cal A}_I(-,-,+,+) &=4(\ap)^{\beta} \ \delta(r-\bar r)\ \theta(r-1)\
\Bigg(\prod_{i<j}^4 z_{ij}^{{h\over 3}-h_i-h_j}
\zbar_{ij}^{{\bar h\over 3}-\bar h_i-\bar h_j}\Bigg)
\cr &\times\, r^{5-\beta\over 3}\,(r-1)^{2-\beta\over 3}\ I(r,\beta) \ .}}

In order to compute the energy integral \ibeta, it is convenient to use the explicit integral representation of the beta function that enters the formfactor in \formf:
\eqn\bett{B(-rw,1+w)=\int_0^1 dx\ x^{-1-rw}\ (1-x)^w\ .}
With $w>0$, this representation is valid for $r<0$ only, {\it i.e}.\ outside our kinematic domain. Nevertheless, we will use it and perform analytic continuation to $r>1$ at the end.
After switching the orders of integration in \ibeta, we obtain:
\eqn\ibetaa{I(r,\beta)=-\Gamma(1-\beta)\ {r\over 2}\ \int_0^1{dx\over x}\ \lf[r\ln x-\ln(1-x)\ri]^{\beta-1}\ .}
In order to compute this integral, we use the binomial expansion:
\eqn\bexp{
\lf[r\ln x-\ln(1-x)\ri]^{\beta-1}=
\sum_{k=0}^\infty\fc{\Gamma(k+1-\beta)}{\Gamma(k+1)\Gamma(1-\beta)}\ (r\ln x)^{\beta-k-1}\ \ln^k(1-x)\ .}
The first term yields the same delta function \jzeroq\ as in Yang-Mills theory,
\eqn\ibetab{{1\over 2}\int_0^1{dx\over x}\ (-\ln x)^{\beta-1}
=2\pi\,\delta\bigg(\sum_{n=1}^4\lambda_n\bigg)\ ,}
while the subsequent terms involve  polylogarithmic integrals:
\eqn\niela{S(\beta-k,k):= {(-1)^{\beta-1}\over\Gamma(\beta-k)\Gamma(k+1)}
\int_0^1{dx\over x}\ (\ln x)^{\beta-k-1}\ \ln^k(1-x).}
In this way, the energy integral \ibeta\ becomes:
\eqn\ibetac{I(r,\beta)
=2\pi\,\delta\bigg(\sum_{n=1}^4\lambda_n\bigg)+{{1\over 2}\ \Gamma(\beta)\ \Gamma(1-\beta)\ (-r)^\beta}\ \sum_{k=1}^\infty (-r)^{-k}\ S(\beta-k,k)\ .}
The expansion coefficients \niela\ are related to
Nielsen's polylogarithm functions $S_{n,k}(t)$~\Kolbig
\eqn\Nielsen{
S_{n,k}(t)=\fc{(-1)^{n+k-1}}{(n-1)!k!}\ \int_0^1\fc{dx}{x}\ \ln^{n-1} x\ \ln^k(1-xt)\ \ \ ,\ \ \ t\in\IC\ ,}
labeled by positive integers $n$ and $k$. Assuming that this function can be extended to complex $n$, $S(\beta-k,k)=S_{\beta-k,k}(1)$. {}This is not difficult for the first term because $S_{n,1}(t)=Li_{n+1}(t)$, where $Li_{n+1}$ is the standard polylogarithm of order $n+1$. Since
$Li_{n+1}(1)=\zeta(n+1)$, $S(\beta-1,1)=\zeta(\beta)$, which can be checked by an explicit computation of the integral \niela:
\eqn\EXPCHECK{\eqalign{
S(\beta-1,1)&=-\Gamma(\beta-1)^{-1}\int_0^1\fc{dx}{x}\ (-\ln x)^{\beta-2}\ \ln(1-x)\cr
&=\Gamma(\beta-1)^{-1}
\sum\limits_{n=1}^\infty\int_0^1
\fc{dx}{x}\;(-\ln x)^{\beta-2}\;\fc{x^n}{n}=\sum\limits_{n=1}^\infty n^{-\beta}=\zeta(\beta)\ ,\ \re(\beta)>1\ .}}
{}For $k>1$,
$S_{n,k}(1)=\zeta(n+1,\{1\}^{k-1})$ \doubref\KolbigZZA\Adamchik, where
\eqn\mzetas{\zeta(n+1,\{1\}^{k-1})=\zeta(n+1,\underbrace{1,\dots,1}_{k-1})=\sum_{n_1>n_2>\dots> n_k}{1\over n_1^{n+1}n_2\cdots n_k}}
is a multiple zeta value (MZV) of depth $k$. At the end, we obtain:
\eqn\ibetad{\eqalign{ {I}(r,\beta)
=&~2\pi\,\delta\bigg(\sum_{n=1}^4\lambda_n\bigg)\cr &+{i\pi\over 2}(-r)^{\beta-1}\sinh\bigg({1\over 2}\sum_{n=1}^4\lambda_n\bigg)^{-1} \sum_{k=0}^\infty (-r)^{-k}\zeta\bigg(-{i\over 2}\sum_{n=1}^4\lambda_n-k,\{1\}^{k}\bigg)\ .}}

The final result, summarized in Eqs. \cglus\ and \ibetad, has  some interesting features. The delta function part of the string amplitude agrees with Yang-Mills theory. The remaining terms should be hence interpreted as ``string corrections'' due to massive string modes, and we will show below that this is indeed the case. They do not come, however in the usual form of an expansion in the string parameter $\alpha'$. In fact, $\ap$-dependence is limited to an overall $(\ap)^\beta$ factor in  \cglus, just to provide the right dimensions. Instead of an $\ap$ expansion, we obtain a small scattering angle expansion in $r^{-1}=\sin^2({\theta\over 2})$ \mand. In traditional string amplitudes, quantum field theory is recovered in the $\ap\to 0$ limit. Here, in celestial amplitudes, quantum field theory is recovered in the kinematic limit of large $r$, that is in the limit of forward scattering at $\theta=0$. Indeed, in this limit, the process is dominated by the exchanges of massless particles.

There is an alternative expression for $I(\beta,r)$ of \ibetad\ which displays the connection to massive string states. The Nielsen polylogarithm function \Nielsen\ can be expanded as \Adamchik\
\eqn\Adam{
S_{n,k}(t)=\sum_{m=k}^\infty\lf[m\atop k\ri]\fc{t^m}{m!\ m^n}\ ,}
where  $\lf[m\atop k\ri]$ are the unsigned Stirling numbers. The relation
\eqn\biop{\sum_{k=1}^m\lf[m\atop k\ri]x^k=\fc{\Gamma(x+m)}{\Gamma(x)}}
allows rewriting
\eqn\arrivee{
 \sum_{k=1}^\infty (-r)^{-k}\ S_{\beta-k,k}= \sum_{n=1}^\infty \fc{1}{n!\ n^{\beta}}\ \fc{\Gamma\lf(-\fc{n}{r}+n\ri)}{\Gamma\lf(-\fc{n}{r}\ri)}\ ,}
After inserting it in \ibetac, we obtain an alternative expression:
\eqn\ibetae{\eqalign{ {I}(r,\beta)
=&~2\pi\,\delta\bigg(\sum_{m=1}^4\lambda_m\bigg)\cr &+{i\pi\over 2}(-r)^{\beta}\sinh\bigg({1\over 2}\sum_{m=1}^4\lambda_m\bigg)^{-1}\sum_{n=1}^\infty \fc{1}{n!\ n^{\beta}}\ \fc{\Gamma\lf(-\fc{n}{r}+n\ri)}{\Gamma\lf(-\fc{n}{r}\ri)}\ .}}
The above form of the energy integral allows identifying the contributions of all mass levels.
In fact, the $n$--th term of the sum originates from mass $\sqrt{n/\ap}$ string excitations. The best way to see this is by converting the energy integral \ibeta\ into a complex integral over the Hankel contour. Then
\eqn\ibetag{ {I}(r,\beta)
=2\pi\,\delta\bigg(\sum_{m=1}^4\lambda_m\bigg)+{i\pi\over (1-e^{-2\pi i\beta})}\ \sum_{n=1}^\infty{\rm Res}_{s=n}\bigg\{\Big({s\over r}\Big)^{-\beta}\! B\Big( {-}s,1+{s\over r}\Big)\bigg\}\ ,}
where the delta function originates from the segment encircling $w=0$ while the residues are due to massive string poles at mass levels $\sqrt{n/\ap}$. Since
\eqn\resi{{\rm Res}_{s=n}\bigg\{B\Big({-}s,1+{s\over r}\Big)\bigg\}=\fc{\Gamma\lf(-\fc{n}{r}+n\ri)}{n!\ \Gamma\lf(-\fc{n}{r}\ri)}\ ,}
Eq. \ibetag\ reproduces Eq. \ibetae.

\newsec{World--sheet as  celestial sphere}

In the previous section, we stressed that the ultra--soft high energy behavior of string formfactors ensures the convergence of energy integrals. The asymptotic form of the four-gluon open string formfactor was exhibited in \formfat\ by using Stirling's formula. It is known that this behavior can be also obtained by using the steepest descent (saddle point) method \GrossAR. Recall that the beta function appears in \formf\ as a result of integrating one vertex position $x$ over the boundary of string disk worldsheet.\foot{Three remaining vertex positions are fixed by $SL(2,\IR)$ M\"{o}bius invariance.} For $s<0$ and $u=-as<0~( a=r^{-1}<0)$,
\eqn\betc{B(-s,1-u)=\int_0^1x^{-1-s}(1-x)^{as}}
The range $(0,1)$ of integration is correlated with one particular color (Chan-Paton) factor. In order to discuss the $s\to -\infty$ limit, one writes
\eqn\betd{ B(-s,1-u)=\int_0^1 x^{-1}e^{-sf(x)}dx~,\qquad f(x)=\ln x-a\ln(1-x)}
and solves the stationary point equation
\eqn\bete{f'(x_0)=0~\Rightarrow~x_0={1\over 1-a}.}
Note that for $a<0$, $x_0$ is on the integration path, where the function reaches the maximum value $f(x_0)=-(1-a)\ln(1-a)-a\ln(-a)$. After applying Laplace's formula, we obtain
\eqn\formfau{F_I(s,u)\sim\sqrt{2\pi as\over(1-a)}
\  (-a)^{as}\ (1-a)^{(1-a)s}}
which is exponentially suppressed at $s\to-\infty$ with $a<0$. The same result follows by applying Stirling's formula. It should be kept in mind that $s<0$ and $a=r^{-1}=\sin^2(\theta/2)<0$ are in the unphysical domain of imaginary center of mass energy and imaginary scattering angle. Note, however, that the stationary point equation ties the world-sheet vertex position $x_0$ to the kinematic cross ratio $r$, identifying a point on the worldsheet with a point on celestial sphere (modulo $SL(2,\IR)$--transformation).

In the physical range of $s>0$, $u<0$, the asymptotic behavior \formfas\ is slightly different from \formfau. If one naively extrapolates \formfau\ to the physical domain,  one will miss the ratio of sines containing the poles of massive string states. The reason is that the integral representation \betc\ is not valid for $s>0$. The integration path needs to be modified to the complex contour known as Pochhamer contour, although for $u<0$ it is sufficient to use any contour running through $x=1$ and circling 0\foot{It is an open contour because it returns to 1 on a different Riemann sheet.}. As far as the asymptotic behavior is concerned, however, the steepest descent method can be used again, with the same stationary point equation \bete\ that yields $x_0=(1-a)^{-1}$, identifying $x_0$ as a point on celestial sphere. Since now $a>0$, $x_0>1$ and  the contour runs twice through the stationary point on two Riemann sheets, on its way out from and back to $x=1$. It is not difficult to see that these two contributions combine to the ratio of sines, as in \formfas.

In celestial amplitudes \melt, the energy dependence is integrated out through Mellin transforms. Is there any limit in which  vertex positions are tied to celestial sphere?
We have already shown that $r\to\infty$ corresponds to the limit of low-energy massless theory. Now we will show that the equivalent of high-energy ``super-Planckian'' limit is reached at $\lambda\equiv\sum_{n=1}^4\lambda_n\to \infty$.
To see this, we rewrite \ibetaa\ as:
\eqn\ibetaf{I(r,\beta)={1\over 2}(-a)^{-\beta}\ \Gamma(1-\beta)\int_0^1 x^{-1}e^{(\beta-1) g(x)}dx~,\qquad g(x)=\ln[-\ln x+a\ln(1-x)]\ .}
Since $\beta=-i\lambda/2$, in order to discuss the limit of $\lambda\to\infty$ we can use the steepest descent method again, now solving the stationary phase equation:
\eqn\betf{g'(x_0)=0~\Rightarrow~x_0={1\over 1-a}.}
The stationary phase point is exactly at the same position as the saddle point \bete\ of the string formfactor. We find
\eqn\ibetag{I(r,\beta)\sim\Gamma(1-\beta) (-a)^{-\beta}\sqrt{\pi a\over \lambda(a-1)}\ln^{\beta-{1\over 2}}[(-a)^a(1-a)^{(1-a)}]}
We conclude that  the string world sheet becomes celestial in the limit of $\lambda=\sum_{n=1}^4\lambda_n\to\infty$. It would be very interesting to establish a relation between the two underlying CFTs.

\newsec{Four--gluon amplitudes in heterotic superstring theory}

In heterotic superstring theory, similarly to type I, the four-gluon amplitude
is related to the Yang-Mills amplitude  \ymhv\ by a simple rescaling,
\eqn\hmhv{{\cal M}_H(-,-,+,+)={\cal M}(-,-,+,+)\ F_H(s,u)\ ,}
with the heterotic formfactor \GrossRR
\eqn\hformf{F_H(s,u)=-{\Gamma(-\ap s_{12})\ \Gamma(\ap s_{23})\ \Gamma(\ap s_{31})\over
\Gamma(\ap s_{12})\ \Gamma(-\ap s_{23})\ \Gamma(-\ap s_{31})}
=-{\Gamma(-s)\ \Gamma(-t)\ \Gamma(-u)\over
\Gamma(s)\ \Gamma(t)\ \Gamma(u)}\ ,}
where $s=\ap s_{12}$, $t=-(1-a)s$, $u=-as$, with $a=r^{-1}$ of \eqq \rat. In the physical domain, $s>0,~t<0,~u<0$, $a\in(0,1)$.
At $s\to\infty$,
\eqn\hformfat{F_H(s,u)\sim 2\ {\sin(\pi as)\sin[\pi(1-a)s]\over\sin(\pi s)}
\times a^{2as}(1-a)^{2(1-a)s}\ ,}
{\it i.e}.\ the formfactor is exponentially suppressed, except at the singular points associated to massive string modes propagating in the $s$--channel.

Recall that the formfactor appears as a result of integrating one of four vertex position over the closed string world-sheet - the Riemann sphere which is usually mapped into a complex plane.\foot{Here again, three vertex positions are fixed by $SL(2,\IC)$ symmetry of the sphere.} It originates from the following complex integral:
\eqn\formH{
F_H(s,u)=-{s\over \pi}
\int_\IC d^2z\ |z|^{-2s-2}\ |1-z|^{-2u}\ (1-z)^{-1}\ .}
Note that this integral converges for $s<0,~u<0$ only, while \eqq \hformf\ represents its analytic continuation to all complex $s$ and $u$.

The computation of heterotic celestial amplitude corresponding to \hmhv\ proceeds in the same way as in the open string case. The amplitude $\tilde{\cal A}_H$ can be cast in the same form as \cglus, but now with $I(r,\beta)$ replaced by
\eqn\hibeta{H(r,\beta)\equiv{1\over 2}\int_0^\infty  w^{-\beta-1}F_H(rw,-w)\,dw~,\qquad\beta\equiv -{i\over 2}
\sum_{n=1}^4\lambda_n\ .}
We use the integral representation \formH\ and,
after switching the orders of integration, we obtain:
\eqn\hbetaa{H(r,\beta)=-\Gamma(1-\beta)\ {r\over 2\pi}\ \int_\IC{d^2z\over |z|^2(1-z)}\
\lf[r\ln |z|^2-\ln|1-z|^2\ri]^{\beta-1}\ .}
Next, we use the binomial expansion as in \bexp, to generate a series expansion in the powers of $r^{-1}$. The first term contains
the complex analog of \ibetab:
\eqn\DeltaIntegralH{
{1\over 2\pi}\int_\IC{d^2z\over |z|^2(1-z)}\ (-\ln |z|^2)^{\beta-1}=2\pi\ \delta\bigg(\sum_{n=1}^4\lambda_n\bigg)\ .}
To see this, note that in polar coordinates $z=\rho e^{i\phi}$, the
angular integral of \DeltaIntegralH\ becomes
\eqn\rinte{{1\over 2\pi}\int_0^{2\pi} d\phi\ (1-\rho e^{i\phi})^{-1}=\cases{
1,&$0<\rho<1\ ,$\cr
0,&$\rho>1\ ,$}}
leaving
\eqn\rintf{2^{\beta-1}\int_0^1{d\rho\over \rho}\ (-\ln\rho)^{\beta-1}=2\pi\ \delta\bigg(\sum_{n=1}^4\lambda_n\bigg)\ ,}
 which is exactly the same field-theoretical delta function as in \ibetab.
 The remaining terms in the binomial expansion of \hbetaa\ are complex integrals of the form very similar to Nielsen's polylogarithms \Nielsen, with single-valued integrands (without branch points), so it is appropriate to consider them as generalized single-valued Nielsen's polylogarithms:
\eqn\CNielsen{
S^{\bf c}_{n,k}(t)\equiv \fc{(-1)^{n+k-1}}{\pi(n-1)!\,k!}\
\int_\IC \fc{d^2z}{|z|^2}\ (1-z)^{-1}\ \ln^{n-1}|z|^2\ \ln^k
|1-zt|^2\  .}
We also define:
\eqn\gcdef{S^{\bf c}(n,k):= S^{\bf c}_{n,k}(1)\ .}
At the end, we obtain:
\eqn\hibetad{\eqalign{ H(r,\beta)
=&~2\pi\,\delta\bigg(\sum_{n=1}^4\lambda_n\bigg)\cr &+{i\pi\over 2}(-r)^{\beta-1}\sinh\bigg({1\over 2}\sum_{n=1}^4\lambda_n\bigg)^{-1} \sum_{k=0}^\infty (-r)^{-k}\ S^{\bf c}\bigg(-{i\over 2}\sum_{n=1}^4\lambda_n-k-1,k+1\bigg)\ .}}

Again, the integrals \gcdef\ can be determined by an explicit computation, \eg\
\eqn\excomp{
S^{\bf c}(1,\beta)=(-1)^\beta\ [1+(-1)^\beta]\ \zeta(1+\beta)\ \ \ ,\ \ \ \re(\beta)>-1\ ,}
which may be computed by using Gegenbauer decomposition, \cf  Appendix \appB.
Furthermore, for the $k=0$--term in the sum \hibetad\ we have\foot{To evaluate this integral we have introduced polar coordinates $z=\rho e^{i\phi}$ and used the integral $\fc{1}{\pi}\int_0^{2\pi}d\phi\ \fc{\cos(n\phi)}{1-\rho e^{i\phi}}=
\rho^n,$\ $0<\rho<1$.}:
\eqn\EXCOMP{\eqalign{
S^{\bf c}(\beta-1,1)&=-\Gamma(\beta-1)^{-1}\pi^{-1}\int_\IC\fc{d^2z}{|z|^2}\ \fc{(-\ln |z|^2)^{\beta-2}}{1-z}\
\ln|1-z|^2\cr
&=[1-(-1)^\beta]\ \Gamma(\beta-1)^{-1}\pi^{-1}\
 \int\limits_{|z|<1}\fc{d^2z}{|z|^2}\ \fc{(-\ln |z|^2)^{\beta-2}}{1-z}\ \lf\{\sum^\infty_{n=1}\lf(\fc{z^n}{n}+\fc{\ov z^n}{n}\ri)\ri\}\cr
&=[1-(-1)^\beta]\ \zeta(\beta)\ \ \ ,\ \ \ \re(\beta)>1\ .}}

Let us compare the above result \hibetad\ with its type I open superstring analogue~\ibetad. The starting points for both expressions are the string formfactors $F_I$ of \eqq \formf\ and~$F_H$ of \eqq \hformf. These are related by the single--valued projection\foot{For a detailed account on this projection we refer the reader to Appendix \appA.}  \StiebergerHBA:
\eqn\singelvalued{
F_H(s,u)=\sv\ F_I(s,u)\ .}
Recall that this relation holds at the level of $\ap$--expansions, which are expansions in the powers of $s$ and $u$. In the next step, these functions are integrated as in Eqs. \ibeta\ and \hibeta. We expect that, at least in some region of parameters $r$ and $\beta$, the relation \singelvalued\ survives Mellin transformations:
\eqn\Singlevalued{
H(r,\beta)=\sv\ I(r,\beta)\ .}
{}For the leading string correction, this entails
\eqn\statement{
S^{\bf c}(\beta-1,1)=\sv\ \zeta(\beta)\ ,}
and more generally:
\eqn\statements{
S^{\bf c}(\beta-k-1,1+k)=\sv\ \zeta\lf(\beta-k,\{1\}^k\ri)\ \ \ ,\ \ \ k=1,2,\ldots\ .}
In fact, putting \EXPCHECK\ and \EXCOMP\ together gives rise to
\eqn\GIVERISETO{
S^{\bf c}(\beta-1,1)=[1-(-1)^\beta)]\ S(\beta-1,1)=[1-(-1)^\beta)]\ \zeta(\beta)\ ,}
which for integer $\beta$ (with $\beta>1$) reduces to \statement.
Although for generic $\beta\in\IC$ (with $\bet\neq k+1$) it may be difficult to give a rigorous proof of the relation \statements, we will present some more supporting arguments in the Appendix \appB.

In any case, while Nielsen's polylogarithms \Nielsen\ seem to be the natural objects for describing open string amplitudes on celestial sphere (\cf  \ibetad), their single--valued version \CNielsen\ appear in closed string amplitudes ({\it cf}.\ \hibetad). This is reminiscent of the periods $S_{n,p}$ appearing in the $\ap$--expansion of the open string form factor \formf\ through the relation~\Kolbig
\eqn\zusammenhang{
S_{n,p}=-\fc{1}{(n-1)!p!}\ \lf.\fc{\p^{n+p-1}}{\p s^{n-1}\p u^p}\ \fc{1}{s}\ F_I(s,u)\ri|_{s=u=0}\ ,}
which in turn as consequence of \singelvalued\ implies for the periods \gcdef
\eqn\Zusammenhang{
S^{\bf c}(n,p)=-\fc{1}{(n-1)!p!}\ \lf.\fc{\p^{n+p-1}}{\p s^{n-1}\p u^p}\ \fc{1}{s}\ F_H(s,u)\ri|_{s=u=0}\ ,}
with the formfactor $F_H$ given in \hformf, respectively.

\newsec{Mixed gauge--gravitational amplitudes}

We proceed to mixed amplitudes involving gauge bosons and gravitons.
In Einstein--Yang--Mills (EYM) quantum field theory, these scattering processes are due
to minimal couplings of gauge bosons to gravitons.
The simplest amplitude involving one graviton and three gluons
\StiebergerLNG\ is
\eqn\eyma{{\cal M}(--,-,+,+)={\langle 12\rangle^4\over\langle 23\rangle\langle 34\rangle\langle 42\rangle}={\omega_1^2\omega_2\over\omega_3\omega_4}{z_{12}^4\over z_{23}z_{34}z_{42}}=r\,{z_{12}\zbar_{34}^2z_{14}\over\zbar_{12}z_{34}\zbar_{13}}\ \omega_4\ .}
In order to obtain the corresponding celestial amplitude, we apply the Mellin transform as in \melt. The result is:
\eqn\ceym{\eqalign{ \tilde{\cal A}(--,-,+,+) &=4\ \delta(r-\bar r)\,\bigg({\zbar_{34}\over z_{12}}\bigg)^{i\lambda_1}\bigg({z_{34}\over \zbar_{12}}\bigg)^{i\lambda_2}\  \bigg({z_{24}\over \zbar_{13}}\bigg)^{i(\lambda_1+\lambda_3)}\bigg({\zbar_{14}\over z_{23}}\bigg)^{i(\lambda_2+\lambda_3)}\cr
&\times\theta(r-1)\,r^3{ z_{14}\,\zbar_{34}\over\zbar_{12}^2\, z_{34}^2\,\zbar_{13}}\, J_{E1}\ ,}}
with the energy integral:
\eqn\jzerone{J_{E1}=\int_0^\infty \omega_4^{i(\lambda_1+\lambda_2+\lambda_3+\lambda_4)}d\omega_4\ .}
The above integral is linearly divergent in the ultra-violet. Under conformal $SL(2,\IC)$, it transforms in the same way as $\omega_4^{1+i\lambda}$, where $\lambda=\sum_{n=1}^4\lambda_n$. Taking this into account, it is easy to see that the amplitude \ceym\ has the transformation properties
 of a four-point correlation function of primary conformal fields with weights
\eqn\weighte{\eqalign{&h_1=-{1\over 2}+{i\over 2}\lambda_1\qquad~\hb_1={3\over 2}+{i\over 2}\lambda_1,\cr&h_2={i\over 2}\lambda_2,\qquad\qquad~~ \hb_2=1+{i\over 2}\lambda_2,\cr
&h_3=1+{i\over 2}\lambda_3,\qquad~~~ \hb_3={i\over 2}\lambda_3,\cr
&h_4=1+{i\over 2}\lambda_4,\qquad ~~~\hb_4={i\over 2}\lambda_4,}}
in agreement with $\Delta_n=1+i\lambda_n$, $J_1=-2$, $J_2=-1$ and $J_3=J_4=+1$.

The amplitude with one graviton and three gauge bosons is also present in heterotic superstring theory, where it incorporates the effects of all massive closed string excitations. Similarly to the case of four gluons, it is related to EYM amplitude by a simple rescaling \refs{\StiebergerLNG,\SchlottererCXA}
\eqn\eymb{{\cal M}_H(--,-,+,+)={\cal M}(--,-,+,+)\ F_H(s,u)\ ,}
where $F_H$ is exactly the same heterotic formfactor \hformf\ as in the pure Yang-Mills case.
This means that its celestial transform has the same form as \ceym, with $J_{E1}$ of \eqq  \jzerone\ replaced by:
\eqn\jhet{J_{H1}=\int_0^\infty \omega_4^{i(\lambda_1+\lambda_2+\lambda_3+\lambda_4)}\ F_H(s,u)\
d\omega_4\ .}
Unlike in quantum field theory, this integral is perfectly convergent
because the string formfactor is exponentially suppressed at high energies, see \hformfat.
The heterotic celestial amplitude can be cast into a form of a four-point CFT correlator:
\eqn\cheym{\eqalign{ \tilde{\cal A}_H(--,-,+,+) &=4(\ap)^{\beta-{1\over 2}} \ \delta(r-\bar r)\ \theta(r-1)\ \Bigg(\prod_{i<j}^4 z_{ij}^{{h\over 3}-h_i-h_j}
\zbar_{ij}^{{\bar h\over 3}-\bar h_i-\bar h_j}\Bigg)
\cr &\times r^{13-2\beta\over 6}\,(r-1)^{1-2\beta\over 6}\ H_1(r,\beta) \ .}}
with:
\eqn\ibeta{
H_1(r,\beta)={1\over 2}\int_0^\infty  w^{-\beta-{1\over 2}}\ F_H(rw,-w)\,dw~,\qquad\beta= -{i\over 2}
\sum_{n=1}^4\lambda_n\ .}
The above integral is the same as the heterotic gauge integral \hibeta\ with the shifted parameter
$\beta\ra\beta -\h$, \ie
\eqn\Shift{
H_1(r,\beta)= H(r,\beta-\half)\ .}
Its expansion in the inverse powers of $r$ has the same form as \hibetad,  although without the field-theoretical delta function because its argument is empty. The first term is of order ${\cal O}(r^{\beta-{3\over 2}})$ and contains $S^{\bf c}(\beta-{3\over 2},1)=[1-(-1)^{\beta-{1\over 2}}]\zeta(\beta-{1\over 2})$.

The amplitudes with two gravitons and two gauge bosons can be discussed in a similar way.

\newsec{Four--graviton amplitudes in heterotic superstring theory}

Over the last thirty years many intriguing connections have been discovered between gauge and gravitations forces. Some of the most interesting ones stem from the classic KLT relations \KawaiXQ\ which allow expressing multi-graviton amplitudes as linear combinations of products of pure gauge amplitudes weighted by kinematic factors. KLT relations were originally derived in the framework of heterotic superstring theory. Since the $\alpha'\to 0$ limit of the heterotic theory is described by a minimally coupled Einstein-Yang-Mills quantum field theory, the graviton amplitudes of Einstein's theory can be expressed in terms of pure Yang-Mills (partial) amplitudes. We are interested in the four-graviton MHV amplitude which can be written as
\eqn\Gravity{
{\cal M}(--,--,++,++)=  {\cal M}(-,-,+,+)\ s_{12}\ {\cal M}'(-,-,+,+)\ ,}
where ${\cal M}(-,-,+,+)$ is the Yang-Mills amplitude of \ymhv\ and the prime indicates $3\leftrightarrow 4$ transposition with respect to the canonical $(1,2,3,4)$ particle ordering. In terms of the energy and celestial coordinates, this becomes:
\eqn\Gravita{
{\cal M}(--,--,++,++)= \omega_4^2\ {|z_{14}|^2|z_{34}|^2\over |z_{13}|^2}\,
\bigg(r\,{z_{12}
\zbar_{34}\over \zbar_{12}z_{34}}\bigg)^2\ .}
The corresponding celestial amplitude is
\eqn\cegr{\eqalign{ \tilde{\cal A}(--,--,++,++) &=4\ \delta(r-\bar r)\ \bigg({\zbar_{34}\over z_{12}}\bigg)^{i\lambda_1} \ \bigg({z_{34}\over \zbar_{12}}\bigg)^{i\lambda_2}\bigg({z_{24}\over \zbar_{13}}\bigg)^{i(\lambda_1+\lambda_3)}\bigg({\zbar_{14}\over z_{23}}\bigg)^{i(\lambda_2+\lambda_3)}\cr &\times\theta(r-1)\,r^4{|z_{14}|^2|z_{34}|^2\over |z_{13}|^2}{ z_{12}\,\zbar_{34}\over\zbar_{12}^3\, z_{34}^3}\, J_{G}\ ,}}
with the gravitational energy integral
\eqn\jzerog{J_{G}=\int_0^\infty \omega_4^{1+i(\lambda_1+\lambda_2+\lambda_3+\lambda_4)}\ d\omega_4\ ,}
which is now quadratically divergent in the ultra-violet.
The above amplitude has the transformation properties of a CFT correlation function of four primary fields with the weights:
\eqn\weightf{\eqalign{&h_1=-{1\over 2}+{i\over 2}\lambda_1\qquad~\hb_1={3\over 2}+{i\over 2}\lambda_1,\cr&
h_2=-{1\over 2}+{i\over 2}\lambda_2\qquad~\hb_2={3\over 2}+{i\over 2}\lambda_2,\cr&
h_3={3\over 2}+{i\over 2}\lambda_3\qquad~~~\,\hb_3=-{1\over 2}+{i\over 2}\lambda_3 \cr
&h_4={3\over 2}+{i\over 2}\lambda_4,\qquad~~\hb_4=-{1\over 2}+{i\over 2}\lambda_4\ .}}

In heterotic superstring theory, four-graviton amplitude is related to Einstein's amplitude by a simple rescaling,
\eqn\Gravitb{
{\cal M}_H(--,--,++,++)= {\cal M}(--,--,++,++)\ F_H(s,u)\ ,}
where $F_H$ is the same formfactor \hformf\ as in the four-gluon case. As a result, we obtain the following celestial amplitude:
\eqn\cegra{\eqalign{ \tilde{\cal A}_H(--,--,++,++) &=4\ (\ap)^{\beta-1}  \ \delta(r-\bar r)\ \theta(r-1)\ \Bigg(\prod_{i<j}^4 z_{ij}^{{h\over 3}-h_i-h_j}
\zbar_{ij}^{{\bar h\over 3}-\bar h_i-\bar h_j}\Bigg)\cr
&\times\, r^{11-\beta\over 3}\,(r-1)^{-1-\beta\over 3}\ G(r,\beta) \ .}}
with:
\eqn\ibethg{
G(r,\beta)={1\over 2}\int_0^\infty  w^{-\beta}\ F_H(rw,-w)\,dw~,\qquad\beta= -{i\over 2}
\sum_{n=1}^4\lambda_n\ .}
Note that the above integrand is square integrable. Once again, we obtain the heterotic integral \hibeta, now shifted by
$\beta\ra\beta -1$, \ie
\eqn\Shiftt{
G(r,\beta)=H(r,\beta-1)\ .}
Its expansion in the inverse powers of $r$ has the same form as \ibetad\ but now,  due to the shift, the field theoretical delta function is absent. The first term contains $S^{\bf c}(\beta-2,1)=[1+(-1)^{\beta}]\ \zeta(\beta-1)$.
The series begins at ${\cal O}(r^{\beta-2})$ order. Note that at large $r$, the $r$--dependent prefactor in \cegra\ grows like $r^{10\over 3}$, therefore four-graviton celestial amplitude blows up in the forward scattering limit of $r\to\infty$.

\newsec{Conclusions}

In this work, we transformed traditional four--point, tree--level scattering amplitudes of gauge bosons and gravitons, describing transitions between their momentum eigenstates, into conformal correlation functions of primary fields on the celestial sphere. In this new representation, the Lorentz group is realized as the conformal symmetry of the sphere. We considered such amplitudes in type I open string theory and in closed heterotic superstring theory, which in the zero slope limit ($\ap\to 0$) are described by minimally coupled Einstein--Yang--Mills systems. This framework ensured ultraviolet finiteness of Mellin transforms of all amplitudes, including the amplitudes involving gravitons that suffer from uncontrollable growth in Einstein's theory while in superstring theory they become supersoft, exponentially suppressed at high energies.

In the momentum space, four--particle amplitudes depend on two kinematic variables that can be chosen to be the center of mass energy and the scattering angle. The scattering angle dependence remains on the celestial sphere as the dependence on one cross--ratio $r$ of the four insertion points of the primary fields; $r$ is constrained to be a real number as expected by the planarity of four--particle scattering processes. On the other hand, particle energies are transformed into the Mellin--dual energy parameters corresponding to the imaginary parts $\lambda_n$ of the dimensions $\Delta_n=1+i\lambda_n$ of primary fields.
It is remarkable that after extracting conformal prefactors, the amplitudes depend only on the total dual energy $\lambda=\sum_{n=1}^4\lambda_n$, so the celestial amplitudes depend on two parameters only, the same number as the usual amplitudes. Actually, this total energy turns out to be zero for pure Yang--Mills amplitudes, as a consequence of the four-dimensional conformal symmetry (scale-invariance) that holds in Yang-Mills theory at the tree--level. There are no such constraints for gravitational amplitudes or for superstring amplitudes, but even in Yang--Mills theory, $\lambda=0$ is not expected to hold beyond the
tree approximation.

One interesting feature of celestial superstring amplitudes is their trivial dependence of the string parameter $\ap$ which is reduced to an overall power factor. In traditional amplitudes,
$\ap\to 0$ is considered as the field theory limit, in which the heavy string modes decouple from the massless spectrum. In celestial amplitudes, all string excitations participate at the same footing. There is, however, a limit in which gauge boson amplitudes approach the Yang-Mills limit. It is $r\to\infty$, which corresponds to  forward scattering, a process dominated by the exchanges of massless particles. In the presence of gravitons, however, this limit is singular.

The celestial superstring amplitudes were presented as series in the inverse powers of $r$, that is as small scattering angle expansions,
with the coefficient determined by the periods (special values) of
Nielsen polylogarithms \niela\ and their complex generalizations \CNielsen.
The latter can be interpreted as single--valued descendants of the real Nielsen polylogarithms.
We explained how the single--valued projection, that relates heterotic and open string amplitudes order by order in their $\ap$ expansions, is implemented at the level of celestial amplitudes.

Another interesting property of celestial superstring amplitudes is revealed in the $\lambda\to\infty$ limit. In this limit, the dominant contributions can be analyzed by using the stationary phase approximation, in a way similar to the saddle point approximation used for extracting the asymptotic ``super--Planckian'' behavior of standard amplitudes. The amplitudes involve integrating over the positions of string vertex operators on the world-sheet, more precisely just one vertex position left after the positions of three remaining ones are fixed by the world--sheet conformal invariance. In the stationary phase approximation, this integral is dominated by the position determined by $r$, therefore the world-sheet can be identified with the celestial sphere. This raises an interesting question whether CFT on celestial sphere is related in some way to free CFT on the world-sheet.

Hopefully, the properties of celestial scattering amplitudes will be helpful in extracting more information about the underlying CFT on celestial sphere.

\vskip 0.75cm
\leftline{\noindent{\bf Acknowledgments}}

\noindent
We are deeply indebted to Andy Strominger for very useful discussions, comments and encouragement. We are also grateful to Laura Donnay, Wei Fan, Angelos Fotopoulos, Andrea Puhm and Shu-Heng Shao for many conversations and communications. We also thank Johannes Br\"odel for helpful discussions and sharing some of his knowledge about multiple polylogarithms.
St. St. thanks Zygmunt Lalak and the Institute of Theoretical Physics of the University of Warsaw for their kind hospitality.
This material is based in part upon work supported by the National Science Foundation
under Grants Number PHY--1620575 and  PHY--1748958.
Any opinions, findings, and conclusions or recommendations
expressed in this material are those of the authors and do not necessarily
reflect the views of the National Science Foundation.


\appendix\appA{The single--valued projection}
\def\sv{{\rm sv}}
\def\SVM{{\zeta^m_{\rm sv}}}
\def\SV{{\zeta_{\rm sv}}}

Power expanding the string formfactors $F_I$ of \eqq \formf\ and~$F_H$ of \eqq \hformf\ w.r.t. small $\ap$
gives:
\eqn\PowerExp{\eqalign{
F_I(s,u)&=1-\zeta(2)\ su-\zeta(3)\ su\ (s+u)
-\fc{2}{5}\ \zeta(2)^2\ su\ (s^2+\fc{1}{4}su+u^2)+\Oc(\ap^5)\ ,\cr
F_H(s,u)&=1-2\ \zeta(3)\ su\ (s+u)+\Oc(\ap^5)\ ,}}
respectively.
Note, that with the relations \zusammenhang\ and \Zusammenhang\ the coefficients
in the powers series expansions \PowerExp\ are given by Nielsen's polylogarithms \Nielsen\ and their complex analogs \CNielsen,
respectively. Moreover, in Ref. \StiebergerWEA\ it has been observed that the second series can be obtained from the first series by applying the following map:
\eqn\ApplyMap{
\sv: \cases{\zeta(2n+1)\mapsto 2\ \zeta(2n+1),& $n\geq1$\ ,\cr
\zeta(2)\mapsto 0\ .&}}
This map represents the single--valued projection $\sv$  introduced in \singelvalued. It is called projection since, \eg $\zeta(2)$--terms are
projected out. More generally, $\sv$ represents a morphism acting on the space of MZVs \mzetas\ mapping the latter to a subspace
of MZVs, namely the single--valued multiple zeta values (SVMZVs) \BrownGIA\
\eqn\trivial{
\SV(n_1,\ldots,n_r)\in\IR\ .}
The numbers \trivial\ can be obtained from the MZVs \mzetas\ by generalizing the map \ApplyMap\ to the full space of MZVs \BrownGIA:
\eqn\mapSV{
\sv: \z(n_1,\ldots,n_r)\mapsto\ \SV( n_1,\ldots,n_r )\ .}
The map \mapSV\ has been constructed\foot{Strictly speaking, the map $\sv$ is defined in the Hopf algebra $\Hc$ of motivic MZVs $\zeta^m$.
In this algebra $\Hc$ the homomorphism $\sv: \Hc\ra\Hc^\sv$, with $\z^m(n_1,\ldots,n_r)\mapsto\SVM(n_1,\ldots,n_r)$ and
$\SVM(2)=0$ can be constructed, \cf \BrownGIA\ for details.} by Brown in Ref. \BrownGIA, where also SVMZVs have been studied from a mathematical
point of view.
For instance, we have $\SV(5,3)=\sv(\zeta(5,3))=-10 \z(3) \z(5)$ and $\SV(7,3)=\sv(\zeta(7,3))=-28\z(3) \z(7)-12\z(5)^2$.

Generically, in the $\ap$--expansion of open superstring tree--level amplitudes the whole space of MZVs \mzetas\
enters  \refs{\GRAV,\SS}, while
closed superstring tree--level amplitudes exhibit only the subset \trivial\ of SVMZVs in their $\ap$--expansion \refs{\GRAV,\StiebergerWEA}.
The relation between open and closed string amplitudes through the the single--valued projection \mapSV\ has been observed in \StiebergerWEA\ and established in \StiebergerHBA.

\appendix\appB{Single--valued Nielsen's polylogarithms}

The Nielsen's generalized polylogarithms \Nielsen\ can be expressed in terms of harmonic polylogarithms $L_w$.
The latter are specified by a word $w$ of letters $e_0$ and $e_1$ as
$$L_{e_0w}(x)=\int_0^x \fc{L_{w}(t)}{t}\ \ \ ,\ \ \ L_{e_1w}(x)=\int_0^x \fc{L_{w}(t)}{1-t}\ ,$$
with
$$L_{e_0^n}(x)=\fc{1}{n!}\ln^n x\ \ \ ,\ \ \ L_{e_1^n}(x)=\fc{(-1)^n}{n!}\ln^n(1-x)\ ,$$
and $L_1=1$ with $1$ being the empty word.
There is also an expression in terms of Goncharov polylogarithms, which are recursively defined through the iterated integral \Gonch
\eqn\goncharov{
G(a_1,\ldots,a_n;t)=\int_0^t\fc{dx_1}{x_1-a_1}\ G(a_2,\ldots,a_n;x_1)\ ,}
In terms of Goncharov polylogs we have:
$$G({\bf 0}_n;x)=L_{e_0^n}(x)\ \ \ G({\bf 1}_n;x)=(-1)^n\ L_{e_1^n}(x)\ .$$
The Nielsen's generalized polylogarithms are expressed as \GDR:
\eqn\asfollows{
S_{n,p}(z)=(-1)^p\ G({\bf 0}_n,{\bf 1}_p;z)=Li_{n+1,1,\ldots,1}(z,1,\ldots,1)\ .}
In particular we have:
\eqn\Asfollows{
G(n,p):=S_{n,p}(1)=(-1)^p\ G({\bf 0}_n,{\bf 1}_p;1)= \zeta(n+1,1,\ldots,1)\ . }
The relation \asfollows\ is derived by first applying shuffle relations as follows:
\eqn\PolyNiels{\eqalign{
S_{n,p}(z)&=(-1)^{n+p-1}\ \int_0^1 \fc{dt}{t}\ G(\underbrace{0,\ldots,0}_{n-1};t) \ G(\underbrace{\fc{1}{z},\ldots,\fc{1}{z}}_{p};t)\cr
&=(-1)^{n+p-1}\ \int_0^1 \fc{dt}{t}\ \sum_{w \in \{0\} \shuffle \{1/z\}} G(w;t)=(-1)^{n+p-1}\ \sum_{w \in \{0\} \shuffle \{1/z\}} G(0,w;1)\cr
&=(-1)^p\ G(\underbrace{0,\ldots,0}_{n},\underbrace{1,\ldots,1}_{p};z)\ .}}
In the last step, shuffle relations as
$G(a_1,a_2,\ldots,a_{l-1},0;z)= G(a_1,a_2,\ldots,a_{l-1};z)G(0;z)-G(a_1,a_2,\ldots,0,a_{l-1};z)\ldots - G(a_1,a_2,\ldots,0,a_{l-2},a_{l-1};z)-\ldots-G(0,a_1,a_2,\ldots,a_{l-1};z)$ are applied to eliminate any term with $a_l=0$
and eventually apply the scaling relation.

Multiple polylogarithms $L_w(z)$ can be combined with their complex conjugates
$L_w(\ov z)$ to remove branch cuts at $0,1$ and $\infty$ and
obtain a single--valued function on $\IC\IP^1\slash\{0,1,\infty\}$.
The single--valued multiple polylogarithms (SVMPS)  have been introduced by Brown \BrownUGM.
They are entirely constructed from holomorphic and anti--holomorphic harmonic polylogarithms as real analytic functions on  the punctured complex plane $\IC\slash\{0,1\}$.
There exists a unique family of SVMPS $\Lc_w(z)$, each of which is a linear combination of the functions $L_{w1}(z)L_{w2}(\ov z)$, with $w_1$ and $w_2$ words in $e_0$ and $e_1$ and they fulfil the
following differential equations
\eqn\DGLH{
\fc{\p}{\p z} \Lc_{e_0w}(z)=\fc{\Lc_w(z)}{z}\ \ \ ,\ \ \ \fc{\p}{\p z} \Lc_{e_1w}(z)=\fc{\Lc_w(z)}{1-z}\ .}
such that   $\Lc_1(z)=1$ and
\eqn\withEx{
\Lc_{e_0^n}(z)=\fc{1}{n!}\ln^n|z|^2\ ,}
and $\lim\limits_{z\ra0}\Lc_w(z)=0$ if $w$ is not of the form $e_0^n$.
Besides, the functions $\Lc_w(z)$ fulfil  the shuffle relations.
Furthermore, we have:
\eqn\WithEx{
\Lc_{e_1^n}(z)=\fc{(-1)^n}{n!}\ln^n|1-z|^2\ .}

In terms of \withEx\ and \WithEx\ the numbers $S^{\bf c}(n,p)$ defined through \CNielsen\ and
\gcdef\ as $S^{\bf c}(n,p)=S_{n,p}^{\bf c}(1)$ read
\eqnn\CNIELSEN{
$$\eqalignno{
S^{\bf c}(n,p)&=\pi^{-1}\ \fc{(-1)^{n+p-1}}{(n-1)!p!}\
\int_\IC \fc{d^2z}{|z|^2}\ (1-z)^{-1}\ \ln^{n-1}|z|^2\ \ln^p|1-z|^2\ \cr
&=\pi^{-1}\ (-1)^{n-1}\
\int_\IC \fc{d^2z}{|z|^2}\ (1-z)^{-1}\ \Lc_{e_0^{n-1}}(z)\ \Lc_{e_1^p}(z)\ ,&\CNIELSEN}$$}
whose integrand is manifestly single--valued.
The complex integral \CNIELSEN\ can be performed by Gegenbauer technique, which uses the expansion \FanUQY
\eqn\Gegenbauer{
\fc{1}{|1-z|^{2\alpha}}=\cases{\sum\limits_{n=0}^\infty C^{(\alpha)}_n(\cos\theta)\ r^n,&$0<r<1$\ ,\cr
\sum\limits_{n=0}^\infty C^{(\alpha)}_n(\cos\theta)\ r^{-n-2\alpha},&$r>1$\ ,}}
with Gegenbauer polynomials $C_n^{(\alpha)}$ and $z=re^{i\theta}$. With the latter method we directly compute, \eg
\eqnn\Directly{
$$\eqalignno{
S^{\bf c}(1,1)&=-\pi^{-1}\ \int_\IC \fc{d^2z}{|1-z|^2\ z}\  \ln|z|^2\cr
&=-\fc{2}{\pi} \int_0^{2\pi}d\theta\ \sum_{n=0}^\infty C_n^{(1)}(\cos\theta)\ e^{-i\theta} \lf(\int_0^1 dr\ r^n\ln r+ \int_1^\infty dr\ r^{-n-2}\ln r\ri)=0\ ,\cr
S^c(2,1)&=\pi^{-1}\ \int_\IC \fc{d^2z}{|1-z|^2\ z}\  \ln|z|^2\ \ln|1-z|^2\cr
&=-\fc{2}{\pi} \fc{\p}{\p\alpha} \int_0^{2\pi}d\theta \sum_{n=0}^\infty C_n^{(\alpha)}(\cos\theta) e^{-i\theta} \lf.\lf(\int_0^1 dr\ r^n\ln r+ \int_1^\infty dr\ r^{-n-2\alpha}\ln r\ri)
\ri|_{\alpha=1}\cr
&=2\ \zeta(3)\ ,}$$}
in agreement with \excomp\ and \EXCOMP, respectively.
In fact, these integrals are just the single--valued projection of the corresponding real Nielsen polylogarithms \Nielsen, i.e. $S^{\bf c}(1,1)=\sv\ S_{1,1}(1)=\sv\ \zeta(2)=0$, and $S^{\bf c}(2,1)=\sv\ S_{2,1}(1)=
\sv\ \zeta(3)=2\zeta(3)$, respectively.
Furthermore, in the same way we easily verify:
\eqn\Moreover{\eqalign{
S^{\bf c}(1,2p-1)&=-\fc{1}{\pi (2p-1)!}\ \int_\IC \fc{d^2z}{|z|^2\ (1-z)}\  (\ln|1-z|^2)^{2p-1}=0\ ,\cr
S^{\bf c}(1,2p)&=\fc{1}{\pi (2p)!}\ \int_\IC \fc{d^2z}{|z|^2\ (1-z)}\  (\ln|1-z|^2)^{2p}\cr
&=2\ \zeta(2p+1)=\sv\ \zeta\lf(2,\{1\}^{2p-1}\ri)\ \ \ ,\ \ \ p\geq 1\ .}}
Alternatively, the complex integrals \CNIELSEN\ may be computed by the residue theorem \SchnetzHQA\ yielding:
\eqn\InGeneral{
S^{\bf c}(n,p)=\sv\ S_{n,p}(1)=\sv\ \zeta\lf(n+1,\{1\}^{p-1}\ri)\ .}
This relation can be proven for integer $n,p\geq 1$ and \statement\ should be read as analytic continuation of the relation \InGeneral.


\listrefs
\end